\documentclass[a4paper,10pt,journal]{IEEEtran}
\usepackage{placeins}
\usepackage{amsmath}
\usepackage{cite}
\usepackage{amssymb}

\usepackage{booktabs}
\usepackage{xcolor}
\usepackage{makecell}
\usepackage{url} 

\usepackage{comment}

\ifCLASSINFOpdf
  \usepackage[pdftex]{graphicx}
  \graphicspath{{../pdf/}{../jpeg/}}
  \DeclareGraphicsExtensions{.pdf,.jpeg,.png}
\else
  \usepackage[dvips]{graphicx}
  \graphicspath{{../eps/}}
  \DeclareGraphicsExtensions{.eps}
\fi

\usepackage{array}
\usepackage{multirow}
\usepackage{longtable}
\usepackage{rotating}
\usepackage{bbding}
\usepackage{amsthm}
\usepackage{subfigure}
\usepackage{bm}
\usepackage{enumitem}
\usepackage{pgfplots}
\pgfplotsset{compat=1.18}

\usepackage{algorithmic}
\usepackage{algorithm}

\begin{document}
%
\title{ESAU: Expeditious Swarm Remote Attestation for Pub-Sub IoT Networks with Untrusted Cloud-based Proxy Verifier}
%
%
%

\author{Ruei-Hau Hsu,~\IEEEmembership{Member,~IEEE,}
        Yu-Hsiang Hu,
        Rafael Kaliski$^{*}$,~\IEEEmembership{Senior Member,~IEEE,} and, Tse-Yuan Yu
    \thanks{This work was partially supported by the National Science and Technology Council (NSTC) of Taiwan under Grants \hbox{112-2221-E-110-023}, \hbox{111-2628-E-110-006-MY3},\hbox{112-2634-F-110-001-MBK}, and \hbox{115-2634-F-110-001} and in part by the Information Security Research Center, National Sun Yat-sen University, Taiwan.}
    \thanks{*Corresponding Author}
    \thanks{The authors are with the Department of Computer Science and Engineering, National Sun Yat-sen University, Taiwan, Emails: rhhsu@mail.cse.nsysu.edu.tw, gp6o413@gmail.com, rkaliski@ieee.org, jas27690@gmail.com}
    
}

%
%

\maketitle

%




\begin{abstract}
As Internet of Things (IoT) devices collect and process data, security is essential. IoT devices often combine with other emerging technologies, such as artificial intelligence (AI) and fifth-generation (5G) communications, to support applications such as smart grids and internet-of-vehicles (IoV). Because IoT devices are deployed remotely, remote attestation (RA) effectively verifies each device's state. However, the scale of IoT deployments makes device management challenging for owners. Therefore, cloud-assisted IoT management services often provide RA solutions. Although cloud-based IoT services can securely manage IoT devices, passive malicious administrators or hacked IoT device owners’ accounts can control IoT devices without cloud IoT service authorization. Therefore, this work proposes a cloud-assisted RA system that addresses three main challenges of IoT device RA: 1) lightweight RA design for resource-limited IoT devices, 2) unauthorized access to IoT devices by passive malicious administrators or compromised owner accounts on the cloud IoT management platform, and 3) RA may not extend to an authenticated key exchange protocol. 
In addition to proposing a swarm RA scheme that addresses these three challenges, this work analyzes the system's security properties and evaluates its feasibility through implementation and simulation.
\end{abstract}

\begin{IEEEkeywords}
Internet of Things, cloud service, remote attestation, key management, security
protocol
\end{IEEEkeywords}

%
\IEEEpeerreviewmaketitle

\section{Introduction}

IoT devices are heterogeneous, ubiquitous, and interconnected. Common IoT deployments occur in smart homes, cities, and factories. 
International Data Corporation (IDC) forecasts that the number of IoT devices will exceed 55.7 billion worldwide by 2025 \cite{01}. 
Due to the massive deployment of IoT devices, manual management and control become challenging. To support these deployments, cloud service providers such as Google, Amazon, and Microsoft offer cloud-assisted IoT services for device management and IoT system security.

Security is a major concern for massive IoT deployments. 
Before IoT devices provide online services, owners must guarantee verifiable device integrity and security. Cloud-assisted IoT services make remote attestation (RA) essential. Yet when the cloud service provider acts as the RA verifier, IoT device security cannot be ensured. Thus, in this work we investigate a swarm RA scheme wherein the device owner does not delegate RA to the cloud-service provider, thus negating concerns about the cloud-service provider's trust.

\subsection{Motivation}
Many IoT devices are lightweight. They can only collect and transmit data because of environmental or cost considerations. These devices lack powerful computing resources to provide comprehensive security mechanisms, making them vulnerable to attacks such as intrusions to steal essential data, insert malicious code, tamper with data, and so forth \cite{03,04}.

In 2016, the domain name system (DNS) provider Dyn suffered a series of distributed denial-of-service (DDoS) attacks \cite{05}. The event disrupted North American and European users' access to internet platforms and services. A botnet \cite{bonet1-gummadi2009not, botnet2-nunes2019vrased} 
Mirai-infected/compromised Internet Protocol (IP) cameras launched the attack. Malicious attacks often profoundly affect IoT applications, such as medical services, industrial facilities, or intelligent vehicles. Therefore, protecting IoT device security is essential.

\begin{figure}[!t]
\centering
\includegraphics[width=\columnwidth]{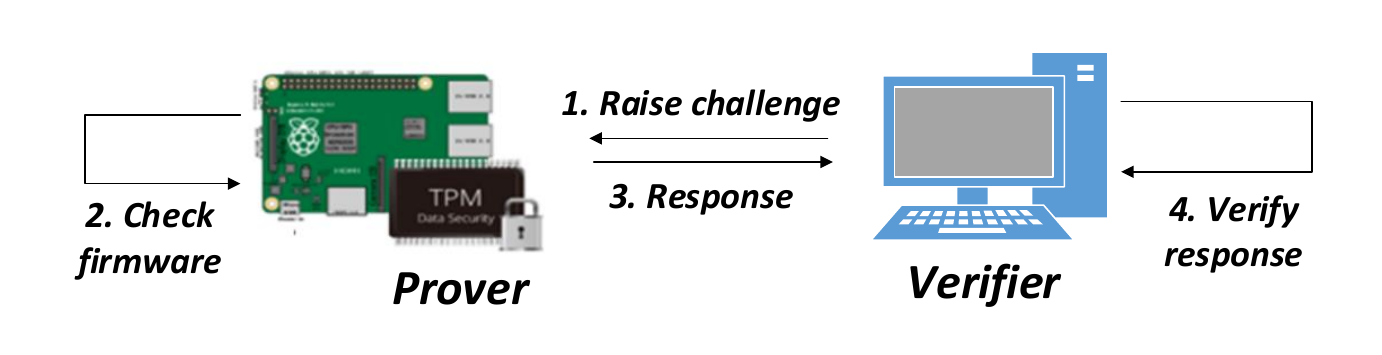}
\caption{General remote attestation process. The verifier uses a challenge-response method to confirm the prover's state. The verifier challenges the prover, which checks its state and generates a response; the verifier then verifies the response.}
\label{fig:General_Remote_Attestation_Process}
\end{figure}

RA \cite{06} is a service mechanism that can remotely prove whether an IoT device is trustworthy. Fig. \ref{fig:General_Remote_Attestation_Process} shows a general RA process consisting of a prover and a verifier. The prover is an untrusted device that must provide evidence to earn the verifier's trust. The verifier is a trusted device that verifies the evidence the prover provides. Thus, the mechanism uses a challenge-response method to confirm the prover's state. During RA, the verifier challenges the prover, and the prover checks the integrity of its configuration (e.g., software) using its trusted detection module. Finally, the verifier receives the response and verifies whether the prover is in a valid configuration.

A traditional RA scheme can attest to IoT devices effectively and securely. However, in an IoT network with many devices, the communication and time costs of traditional RA increase substantially \cite{07}. Recently, many researchers have proposed swarm-based RA schemes \cite{07,08,09,10,11,12}, also called collective remote attestation protocols (CRA), to improve traditional one-to-one RA. CRA allows the verifier to use a unique measurement to check the scale of the IoT network. CRA reduces the communication and computation costs of RA for large-scale IoT networks and improves its scalability.

Many RA studies focus on improving the efficiency and safety of heterogeneous devices in IoT environments. However, these studies assume the commissioned verifier is fully trustworthy. In other words, we cannot ensure the device is safe even after performing RA. Between successful authentication and the user's access to the device for mutual authentication, an adversary can attack during the interval when the user still believes the device is safe. In addition, the owner often delegates integrity verification to a cloud service. Cloud services can reduce the burden of managing many devices. However, the cloud service provider may store device information or use the remaining information to access the device without the owner's authorization. Consequently, cloud services can threaten IoT devices and IoT networks.   

Forbes News reported an example of why you should not fully trust a manufacturer or cloud service provider \cite{15}. They found that a popular mobile phone brand 
may record users' activities, such as private user data and visited-website data, even in supposedly private incognito mode. Moreover, some applications silently access user information and send it back to private company databases. Over time, many more examples of misbehavior by third-party or cloud services have been uncovered. Therefore, it is impractical to trust third-party services fully. Although we believe cloud services are secure, passive malicious administrators may still collect our statistics and use this information to perform unauthorized operations. 

Consequently, device owners need to protect device security and software integrity and manage them efficiently and securely in an environment with many heterogeneous IoT devices.

\subsection{Contributions}
This study proposes the Expeditious Swarm Attestation in Untrusted Environment (ESAU) RA scheme against passive malicious administrators of cloud IoT services, enabling effective management and control of many devices through cloud-assisted IoT management services. In addition, we propose multi-factor authentication to prevent passive malicious administrators from unauthorized access. Furthermore, our proposed scheme can perform swarm RA on target devices selected in cloud-assisted IoT management services, where performing RA to ensure the security of IoT devices and exchanging session keys between the device owner and the devices can be completed simultaneously (the cloud management services act as a proxy verifier for the IoT device owner). The contributions of this work are as follows.


\subsubsection{Effectively manage/attest a large number of IoT devices by cloud IoT service}
The proposed RA system manages many IoT devices through cloud IoT services. It allows device owners to manage their devices by storing critical information and controlling them in the cloud.
\subsubsection{Prevent passive malicious administrators of cloud services from making unauthorized actions}
Through a multi-factor authentication design, the cloud-assisted IoT management service requires each device owner's authorization to perform access control on IoT devices. This design ensures the device owner can manage devices with the cloud IoT platform while preventing cloud IoT platform administrators from accessing devices without authorization.
\subsubsection{Exchange session key during RA}
RA and session key exchange may occur simultaneously, ensuring that the IoT device refreshes the session keys required for secure communication and control each RA period.

\subsection{Organization}
The rest of this work is presented as follows: We first review the background in section \ref{sec:Related_Works}. 
Next, in Section \ref{sec:System_Security_Models}, we describe the system model, threat model, and security requirements. 
Then, in section \ref{sec:Proposed_Protocol}, we describe the details of the proposed protocol. Next, in Section \ref{sec:Sec_Analysis_Compare}, we analyze the system's security features and efficiency. Section \ref{sec:Implement_Perf_Eval} presents and discusses the implementation and performance of our system.
Finally, Section \ref{sec:Conclusion} presents our conclusion.

\section{Background}
\label{sec:Related_Works}
This section introduces cloud IoT services, remote attestation, TPMs, related tools, and authentication codes.
\subsection{Cloud IoT Service}
Most well-known cloud service providers and enterprises now provide secure cloud IoT services, such as Google Cloud IoT, 
Microsoft Azure 
Amazon Web Services (AWS), 
and IBM Watson. 
Several of these companies provide open-source cloud IoT management platforms \cite{20}. The platforms mentioned above make IoT device management more convenient. Thus, many users place their IoT devices on cloud IoT management platforms to improve management efficiency, flexibility, and scalability, and to reduce cost considerations.

However, building a secure cloud platform usually requires meeting many security requirements and defending against corresponding malicious attacks across different device types. Ray et al. \cite{21} surveyed 26 IoT cloud platforms to provide introductions and insights based on various service requirements. Users can choose the appropriate platform based on their needs. Finally, this work discusses issues faced when operating on cloud platforms, such as heterogeneity, IoT node identity, and energy management, and proposes possible improvements. Sajid et al. \cite{22} discussed challenges and security analyses for cloud-assisted IoT-based supervisory control and data acquisition (SCADA) systems in industrial IoT.

\subsection{Remote Attestation}
Generally, depending on the inspection modules, RA can be divided into software-based, hardware-based, and hybrid-based attestation.
\subsubsection{Software-based RA}
Software-based RA usually relies on timing information to let the verifier measure the correctness of the device's firmware, along with a few special hardware requirements. Thus, these methods usually require strong assumptions to ensure the correctness of the device's software/firmware because they have strict timing requirements. In the current IoT network environment, these assumptions must be revised.

In prior work on software-based RA, SWATT: SoftWare-based ATTestation for Embedded Devices (SWATT), proposed by Seshadri et al. \cite{13}, uses the attestation time difference to determine whether the device is infected. Because the compromised device responds to the attestation request with valid results, it must redirect memory access to the memory location where the original code is stored. However, this memory-access redirection adds time overhead and causes the verifier to detect damage. Therefore, the protocol measures the device through this measurement.

Ammar et al. proposed Verify \& Revive: Secure Detection and Recovery of  Compromised Low-end Embedded Devices \cite{40}, which uses a formally verified isolation technique to emulate a Trusted Platform Module (TPM) to achieve hybrid-based RA in low-end devices. However, it is difficult to demonstrate that this approach matches the original hardware security of a TPM.

\subsubsection{Hardware-based RA}
Hardware-based RA aims to address the limitations of software-based RA. It usually equips the device with a hardware detection module \cite{25}, which is a trusted execution environment. In addition to storing the digest of the device's image configuration in secure storage, it can also perform calculations and generate a secure key. The most famous is the TPM \cite{23}.

\subsubsection{Hybrid-based RA}
Although hardware-based RA can provide reliable, secure platforms to ensure device safety, the hardware module usually carries a significant cost burden. Hardware-based RA may not be possible in lightweight IoT devices. Hybrid-based attestation requires only minimal hardware security storage and is currently suitable for IoT environments, such as SMART \cite{27}, TrustLite \cite{28}, TyTAN \cite{29}, SANCUS \cite{30},  and TRAIN\cite{frolikov2025toctou}.

Askan et al. proposed the first swarm attestation scheme called SEDA \cite{07}. The scheme arranges IoT devices into a spanning-tree topology and recursively attests to a leaf. Each device on the path collects and aggregates attestation responses from the leaf devices. Finally, the tree's root device aggregates all responses and sends them to the verifier for verification. 

AmBrosin et al. proposed SANA \cite{08} to improve the computing resources of aggregation from the SEDA scheme. They designed an Optimistic Aggregate Signature (OAS) algorithm to enable IoT devices with low computing resources to perform aggregation actions effectively. However, signature computation costs remain high for low-end devices. Xiong et al. proposed a new network authentication scheme \cite{09} that uses cloud computing to help IoT devices with low computing resources perform report aggregation. Thus, it addresses SANA's high computational cost while protecting device privacy.

Ammar et al. argued that existing collective RA techniques could not effectively solve collective RA for heterogeneous devices because of different attestation requirements and responses. Therefore, they proposed WISE \cite{12}, which aims to optimize communication cost while preserving sufficient security and shows that the scheme can adapt to dynamic heterogeneous IoT networks with high efficiency and scalability.

ESDRA \cite{11}, proposed by Boyu Kuang et al., is the first many-to-one attestation method to improve the single point of failure. This work uses a distributed attestation approach to attest to each device's integrity and uses an accusation mechanism to report compromised nodes. This scheme mutually attests to device states through neighbors and calculates a trust value for each device. Finally, the protocol uses information reported by each neighbor's device and that neighbor's trust value to confirm whether the target device is in a good state.

Ibrahim et al. proposed a scheme, DARPA \cite{10}, that uses an absence-detection approach to detect physical attacks because physical attacks require a non-negligible amount of time. The device shares the heartbeat message with neighbors to ensure the safety of neighbors and itself. If the device does not send a heartbeat to the neighbor within the expiration time, the neighbor reports it to the verifier.

Furthermore, several studies address collective remote attestation for either higher efficiency or specific security requirements~\cite{RA-Petzi2022-USENIXSEC02, RA-PROVE-Dushku23-JISA23, RA-ACFA-Caulfield23-USENIXSEC23, RA-Ebrahimi24-NDSS24}. They either detect cyberattacks that change only the execution order of program code to enable intrusion into target systems or cause damage without changing the status of memory or storage on victim systems, or they consider the efficiency of RA operations with a highly robust management system. Thus, the authors proposed a Hybrid (hardware-mixed-with-software) remote attestation to defend against these attacks, where a dedicated hardware mechanism records function calls in logs by the TPM~\cite{RA-ACFA-Caulfield23-USENIXSEC23}. However, a dedicated procedure is needed to record each instruction's execution. Specific blockchain-based RA schemes also exist for a publisher-subscriber RA model~\cite{RA-PROVE-Dushku23-JISA23, RA-Petzi2022-USENIXSEC02, RA-Ebrahimi24-NDSS24, shen2026ltraa}. Using blockchain for RA to build a service that does not rely on a highly robust system is impressive. However, building a blockchain system requires significant investment, which is unrealistic for an IoT service provider that is not large enough. Thus, this work aims to design a swarm RA with reasonable costs for constructing and managing an RA service. Thus, in the comparison, this work discusses only related works similar to the proposed scheme in terms of costs and management.

\begin{figure}[!t]
\centering
\includegraphics[width=0.8\columnwidth]{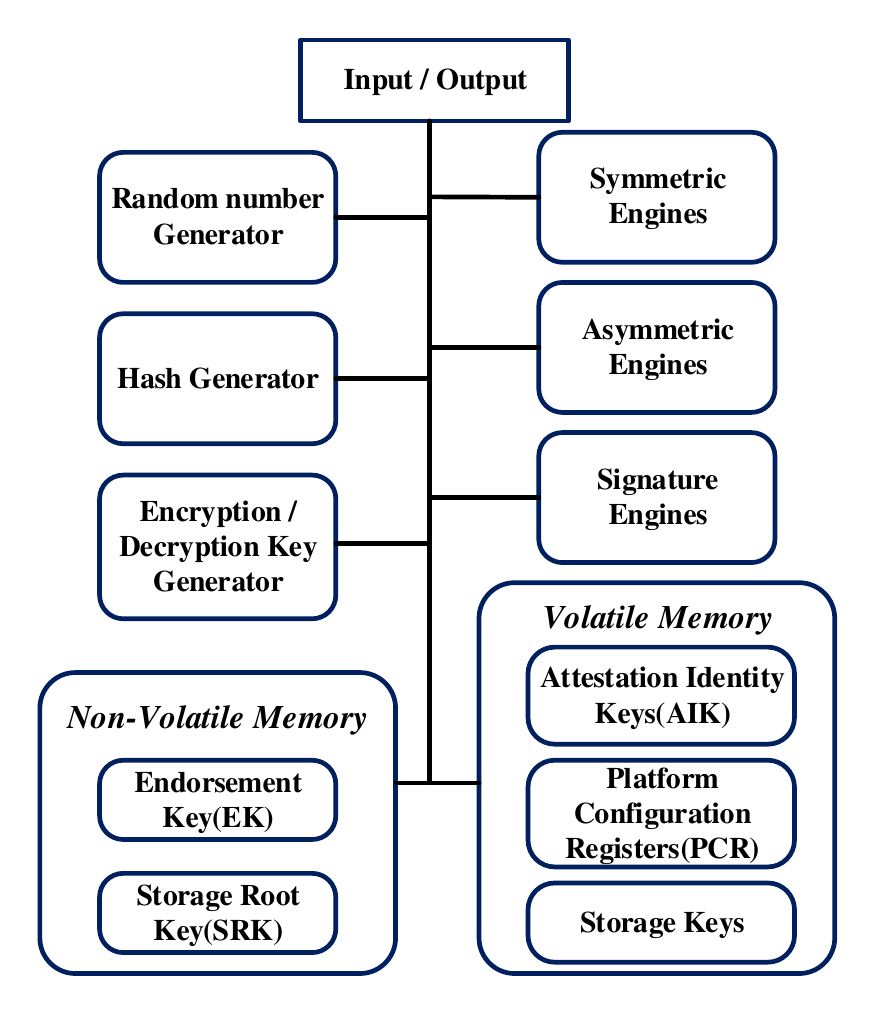}
\caption{TPM architecture and key storage locations. The Endorsement Key is pre-installed, and the TPM generates all other keys.}
\label{fig:TPM_Architecture}
\end{figure}

\subsection{Trusted Platform Module Architecture}
\label{sec:Prelim_TPM}

As shown in Fig.~\ref{fig:TPM_Architecture}, each TPM has a unique pair of endorsement keys (EK) that authenticate the identity of this TPM. The TPM stores it in its secure, tamper-resistant persistent storage, so an operator cannot extract it to prove its identity; the owner can use EK to generate an attestation identity key (AIK) with a third-party privacy certification authority (PCA). The AIK replaces the EK to prove the TPM's identity and sign data the TPM generates.

\begin{figure}[!t]
\centering
\includegraphics[width=\columnwidth]{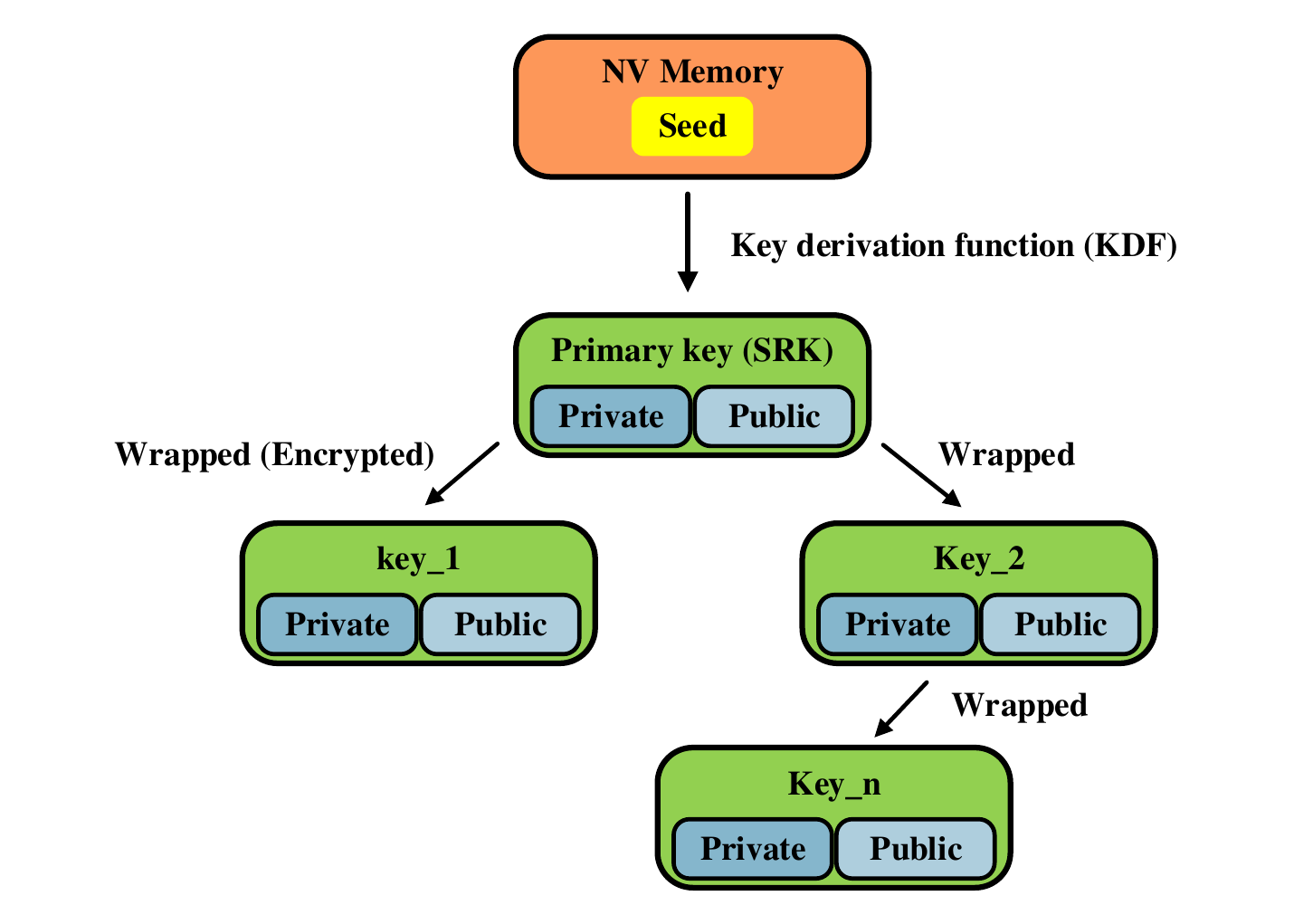}
\caption{The key generation topology of TPM.  The EK and a random seed used to generate the SRK are stored in non-volatile memory. SRK is later used to generate all subsequent keys, such as AIK and application keys. The key generation depends on the parent keys.  
}
\label{fig:key_generation_topology_TPM}
\end{figure}

As shown in Fig. \ref{fig:key_generation_topology_TPM}, the TPM key derivation function (KDF) uses the pre-installed EK and a seed to generate a unique primary key called the Storage Root Key (SRK) \cite{26}, and this key never leaves the TPM or leaks out. When keys leave the TPM, their parent key wraps (encrypts) them. Therefore, the key topology forms a tree structure, with the primary key SRK as the root and the upper keys wrapping branches or leaves.

A TPM contains multiple Platform Configuration Registers (PCRs) that verify the device's integrity through stored metric values; the TPM specification defines at least 24 PCRs. 
PCR values can change when code executes, and the TPM allows only extended operations (extended operations are irreversible because they update the value through a hash function, such as $new value = H(old value||digest)$) or reset operations to change a PCR value. Therefore, the TPM can use sealing/unsealing to bind or store important device information.

The remaining space inside TPM hardware includes other components, such as a random number generator and a cryptographic key generator. They can perform some simple cryptographic actions in a trusted execution environment. Recently, TCG also used the concept of a library 
to create TPM2.0, which improves the flexibility of TPM applications and adds a variety of hash and encryption algorithms.

The Multi-tier randomized remote attestation in IoT networks (MTRA) \cite{14}, proposed by Tan et al., uses a TPM to perform RA on IoT devices with different resources and computing capabilities. They use its sealing/unsealing functions and PCR-reading functions to prove the device's safety to the verifier, which may be another device with a TPM or a base station. They also proposed a key chain method for authentication 
to ensure the freshness of attestation through the irreversible feature of the one-hash function (A new key for the next round of remote attestation is generated after the current remote attestation is finished for a key chain design)
Finally, they proved that MTRA can defend against the Time-Of-Check-To-Time-Of-Use (TOCTTOU) attack, 
where attackers exploit device vulnerabilities after attestation and before the device is used for its intended tasks. In the next two sections, we discuss TPM tools and the Aggregated Message Authentication Codes used by TPMs.

\subsubsection{TPM Tools}  
\begin{figure}[!t]
\centering
\includegraphics[width=1.0\columnwidth]{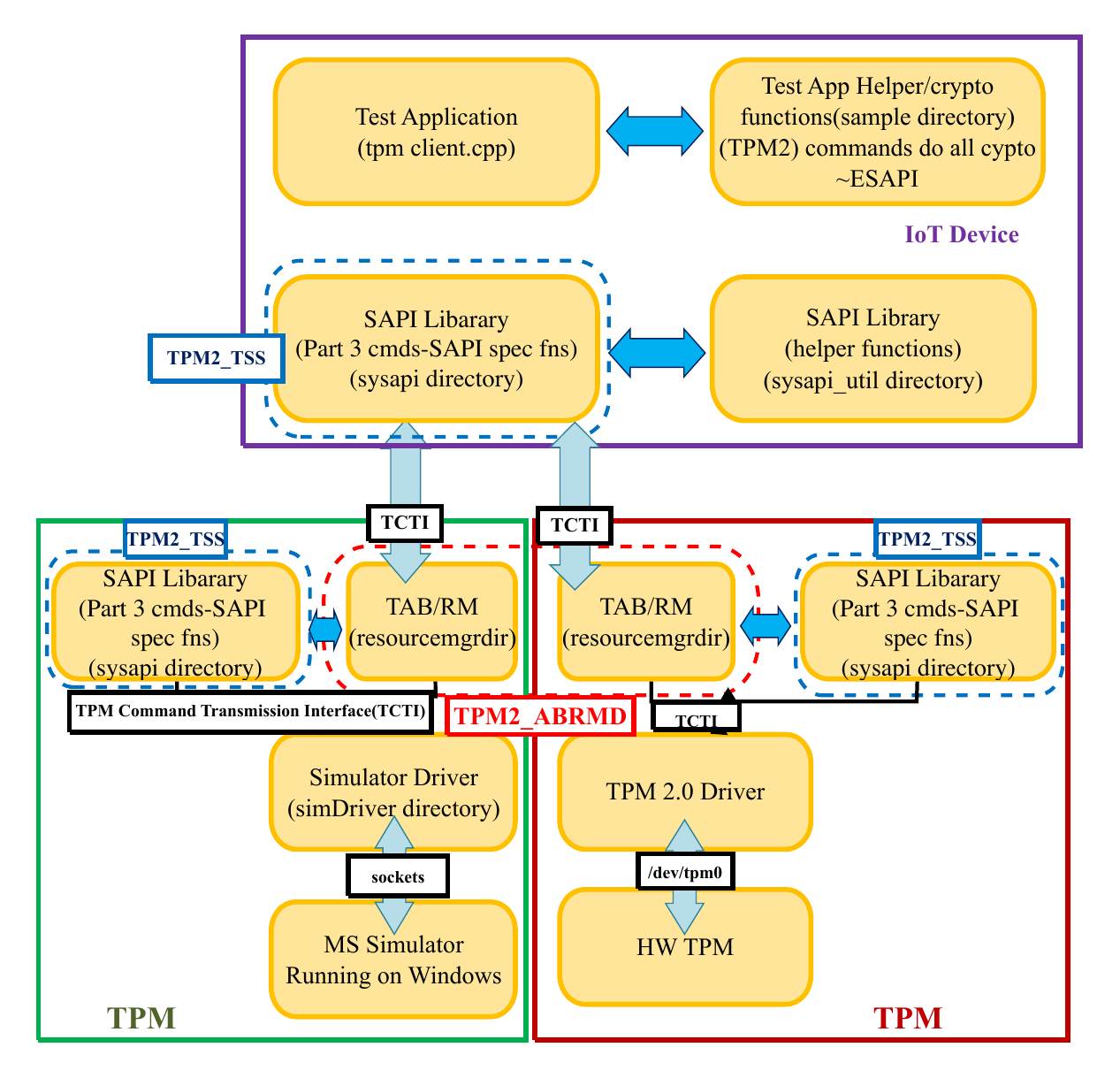}
\caption{The block diagram for using the TPM.  The TPM simulator on the left interacts with the IoT Device during debugging. During deployment, the TPM hardware on the right interacts with the IoT device.}
\label{fig:TPM_Block_Diagram}
\end{figure}

\hfill\\\indent
To operate a TPM, we must issue commands through its firmware. Therefore, we need to run the following three source code modules to establish a connection with the TPM firmware: TPM2 Software Stack (tpm2-tss) \cite{31}, TPM2 access broker \& resource manager (tpm2-abrmd) \cite{32}, and TPM2 tools (tpm2-tools) \cite{33}, as shown in Fig. \ref{fig:TPM_Block_Diagram}.

$\bullet${\bf TPM2-tss:}
TSS is called a trusted software stack. It supports the TPM and provides an interface for the operating system and applications to use it. For this research, the source code implements the software stack in the TPM 2.0 specification defined by TCG.

$\bullet${\bf TPM2-abrmd:}
This system daemon implements the TPM2 access broker (TAB) and resource manager (RM) specification from TCG. 
It sends and receives TPM commands and responses between the client and the server. It is mainly used to protect and manage TPM hardware resource allocation. \\

The rest of this section describes the TPM functions used in the proposed protocols. Some functions require executing multiple TPM commands. First, we introduce basic TPM commands, then explain how each TPM function is executed. For the other functions not mentioned, please refer to the above-mentioned specifications.

$\bullet${\bf tpm2\_create:} This creates a child object, a key, or a sealing object. The proposed protocol creates a keyed Hash-based Message Authentication Code (HMAC) key and seals a secret to the TPM.

$\bullet${\bf tpm2\_pcrread:}
This reads the PCR values and outputs them to the screen. This is important in our protocol; we use the PCR values to determine whether the device is in good status.

$\bullet${\bf tpm2\_unseal:}
This unseals the sealed object in the TPM. This protocol unseals the secret the TPM uses to compute a session key during the attestation phase.\\

$\bullet${\bf tpm2\_getrandom:}
This generates a random number in the TPM. The TPM uses the generated number in further ECDH operations.\\

Our protocol uses some TPM functions built from a series of TPM commands, such as the TPM seal/unseal function. The current PCR value must match the PCR values stored at the time of sealing to pass verification. 

\subsubsection{Aggregate Message Authentication Codes} \hfill\\\indent
According to aggregate message authentication codes~\cite{34} proposed by Katz et al., Message Authentication Codes (MACs) can be aggregated efficiently and safely. The aggregated MAC (AMAC) length equals that of the basic MAC. Tag generation efficiency also matches that of the standard MAC, requiring only $O(n)$ bitwise operations for a $b$-bit output.
In addition, the computational complexity efficiency 
of tag aggregation and verification of aggregated tags grows linearly with the number of tags.


This scheme is referred to as AMAC, where the aggregation of signatures generated by MAC functions is defined as follows \cite{35, 36}.

$\bullet${\bf Message Authentication Code ($\bm{Mac}$) Algorithm:}
To input a key $k \in \{0, 1\}^n$ and message $m \in \{0,1\}^*$, and output the tag $t$ via the standard MAC algorithm per \eqref{eq:Mac}.
\begin{equation}
    \label{eq:Mac}
    tag\ t = Mac(k, m)
\end{equation}

$\bullet${\bf Aggregation ($\bm{Agg}$) Algorithm:}
This algorithm takes the input of two sets of message/id pairs $M^1 = \{(m^1_1, id^1_1), ..., (m^1_i, id^1_i)\}$ and $M^2 = \{(m^2_1, id^2_1), ..., (m^2_j, id^2_j)\}$, where $i$, $j$ are the number of elements in the sets, and the two tags, $t^1$ and $t^2$, of $M^1$ and $M^2$. Then, the algorithm outputs a new tag $t$ and a new set $M = \{(m^1_1, id^1_1), ..., (m^1_i, id^1_i), (m^2_1, id^2_1), ..., (m^2_j, id^2_j)\}$  by using the exclusive-OR (XOR) operation to aggregate the tags via \eqref{eq:Agg}.
\begin{equation}
    \label{eq:Agg}
    new\ tag\ t = tag\ t^1 \oplus tag\ t^2
\end{equation}

$\bullet${\bf Verification ($\bm{Vrfy}$) Algorithm :} 
The input to the $Vrfy$ algorithm is: 1) keys $k_1, ..., k_l \in \{0, 1\}^n$, 2) a set $M = \{(m_1, id_1), ..., (m_l, id_l)\}$ of message/id pairs, where $l$ is the number of all participating elements, and 3) a tag $t$. Then, the algorithm calculates the value $t^\prime = \oplus^l_{j=1}Mac_{k_j}(M_j)$. Finally, if $t^\prime$ equals $t$, verification succeeds, and the message is confirmed. 

\begin{figure}[t]
\centering
\includegraphics[width=1.0\columnwidth,clip,trim=0mm 20mm 0mm 0mm]{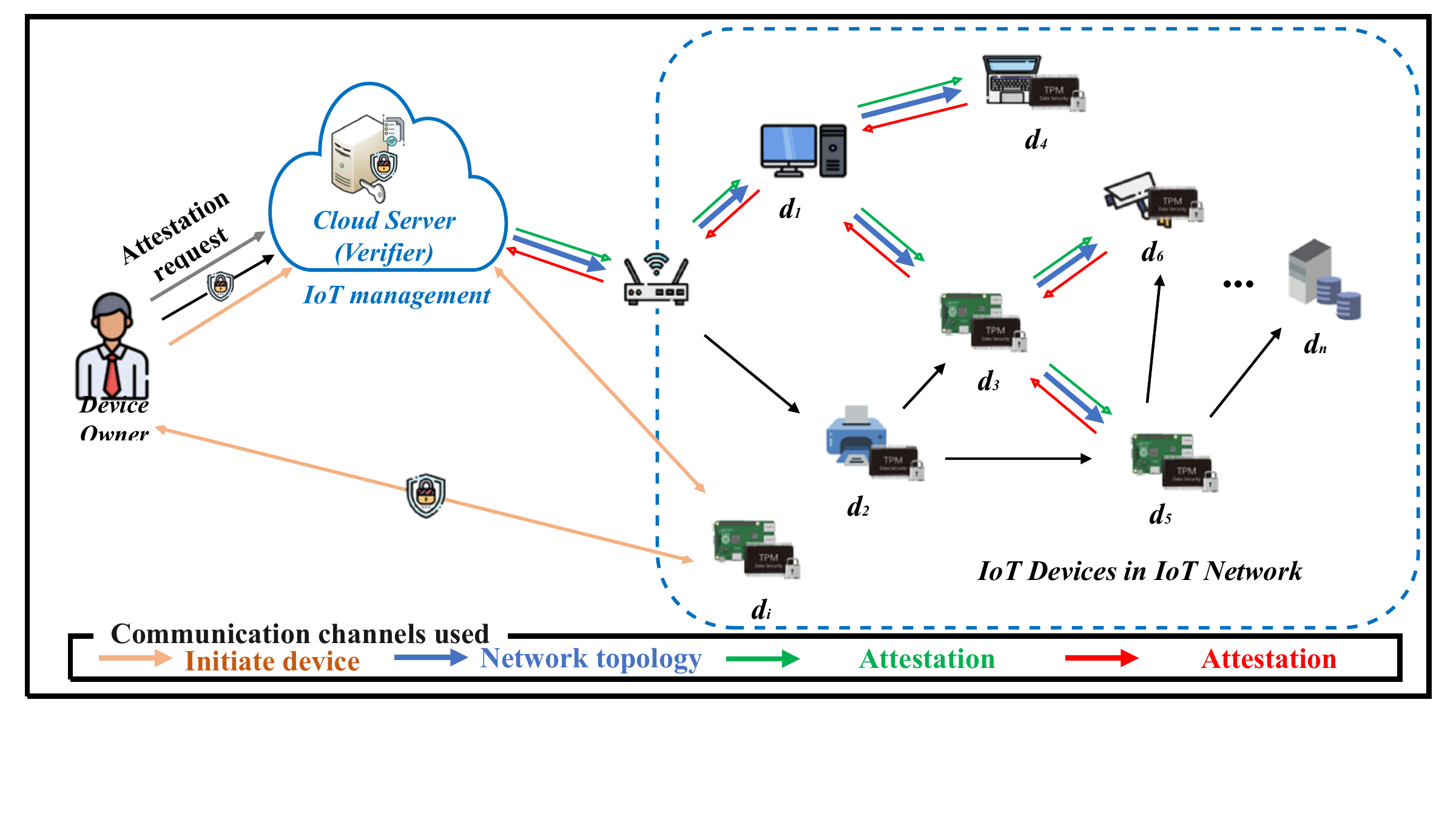}
\caption{Proposed cloud-assisted remote attestation system model: During the Initialize device phase, the device owner initializes the IoT devices directly and then registers them using the cloud-based IoT management service. During the network topology selection phase, the device owner selects IoT devices via the IoT management service. During the attestation challenge phase, the device owner sends a per-device challenge via the IoT management service. During the attestation response phase, each IoT device responds to the device owner via the IoT management service.}
\label{fig:Cloud-assisted_remote_attestation}
\end{figure}

\section{System and Security Models}
\label{sec:System_Security_Models}
This section presents the system and security model to clarify the difference between the proposed swarm-based remote attestation and related works. This section also defines the security against potential attackers in the proposed system model.

\subsection{System Model}
We consider a static IoT network formed by heterogeneous IoT devices, such as in a smart factory or company. The devices can communicate with the network and other devices through their physical IP addresses or internal IP addresses under the router. Moreover, our proposed scheme has three phases: initialization, registration, and attestation.

\subsubsection{System entity}
As shown in Fig. \ref{fig:Cloud-assisted_remote_attestation}, our system model includes three types of entities: the device owner, the cloud-assisted IoT management service, and the IoT device.

{\bf Device owner:}
The device owner can manage all its devices through the cloud-assisted IoT management service and directly access and control the device when necessary. In addition, the device owner needs to remember only two passwords: one for the cloud-assisted IoT management service and another for the TPM in each device.

{\bf Cloud-assisted IoT management service:}
The cloud-assisted IoT management service is a semi-trusted but secure third party with sufficient 
computing resources and is not compromised by attackers, who may conduct malicious operations to disrupt the required security of the system. 
The cloud-assisted IoT management service includes secure servers that store IoT device information and a secure verifier that verifies attestation responses. It can also form a specific network topology in the IoT environment according to the device owner's needs.

{\bf IoT device:}
The IoT device has limited computing resources yet is equipped with a secure module, i.e., a hardware or software TPM, to reduce its vulnerability to malicious attacks.\\

\subsection{Threat Model}
We assume the device owner initializes the IoT device, sets up the TPM in a secure environment, and establishes a secure channel, such as password-based Transport Layer Security (TLS), to log in to the cloud IoT service and communicate securely with the server. We also assume that there are two types of adversaries $\mathcal{A}$.

$\bullet${\bf $\mathcal{A}$ is a software-only attacker.}
$\mathcal{A}$ can use a remote attack to invade the software of the IoT device to steal information about the IoT device or inject malicious programs, but can not physically damage the device's hardware. $\mathcal{A}$ can eavesdrop, insert, and modify all messages except TLS messages that pass through the cryptographic base.

$\bullet${\bf $\mathcal{A}$ is an honest but curious cloud IoT service.} 
It can try to probe the shared secret between the device owner and the device to impersonate a legitimate device owner using the device owner's information on the server.

Finally, we do not consider DoS attacks because such attacks block the receipt of all messages, such as requests, challenges, and responses. However, one of our purposes is to detect whether the IoT device is infected. If it is infected, we can isolate or restrict it as early as possible to prevent it from being used to perform this type of attack on other IoT devices.

\subsection{Security Requirements}
In addition to the threats mentioned above, our proposed collective RA scheme applies to large IoT network environments. Therefore, our scheme needs to conform to the following properties:

$\bullet${\bf Mutual authentication:}
Before using cloud IoT services, all entities involved in the protocols should ensure mutual identification. Each entity must be able to confirm the message from the correct source. In addition, the RA protocol should ensure freshness so that attestation is up to date. The server that initiates attestation should provide a different nonce for each attestation and set a time limit for it.

$\bullet${\bf Authenticated key exchange:}
Before RA, the device owner and the IoT device should identify each other through mutual authentication. In addition, both of them should be able to exchange a session key during RA.

$\bullet${\bf Muti-factor access control:}
The device should ensure that only users with multi-factor keys can access it. It must prevent malicious users from gaining unauthorized access.

$\bullet${\bf Integrity of attestation data:}
The device should ensure that the attestation data is not modifiable. Moreover, the device owner and the server should be able to check the integrity of the attestation data.

$\bullet${\bf Against TOCTTOU attack:}
The scheme should defend against time-gap attacks to prevent malicious programs from attacking the time between device authentication and device access.

\subsection{Security Definitions}
In this section, we introduce some security definitions, such as the following:

\begin{figure*}[!t]
\centering
\includegraphics[width=0.8\textwidth]{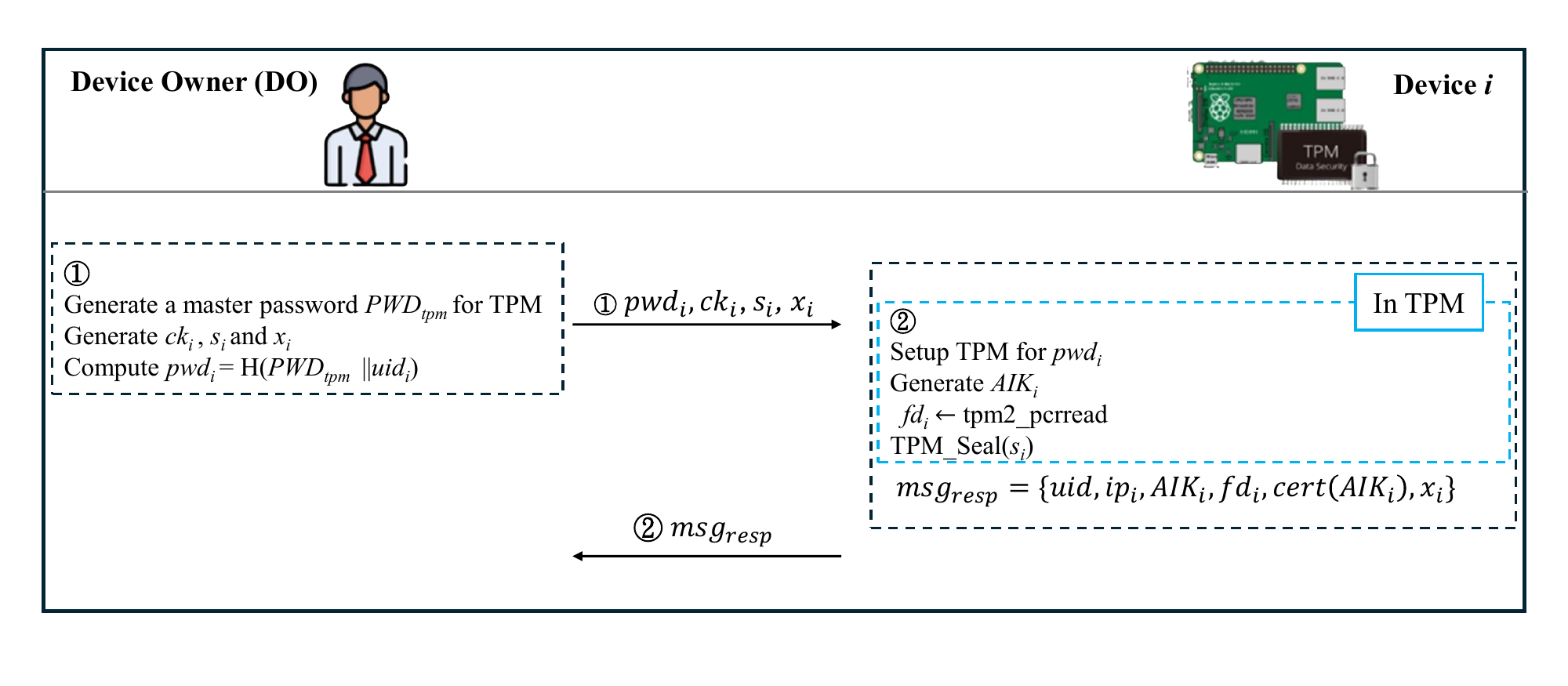}
\caption{Initialization phase: The device owner generates the session key, session secret, and unique TPM password. The device owner then transmits these to the device. After that, the Device's TPM reads the firmware digest, seals the session secret, generates the attestation identity, and responds to the device owner with the device's unique identity.}
\label{fig:Init_Phase}
\end{figure*}

\newtheorem{definition}{Definition}
\begin{definition}
Secure Swarm Attestation

The objective of swarm attestation \cite{07} is for the verifier $V$ to confirm if all devices in a swarm $S$ authenticated by the device owner are performing an attestation. Swarm attestation was demonstrated in the security experiment $Exp_{\mathcal{A}}$, where an adversary $\mathcal{A}$ interacts with devices in $S$ and $V$. $\mathcal{A}$ can tamper with the software of any device $D \in S$ and can eavesdrop and modify the messages between $D \in S$ and $V$. Moreover, $\mathcal{A}$ can also send arbitrary messages to any $D \in S$. After a polynomial number (in terms of the security parameters $\ell_{N}, \ell_{g}, \ell_{mac}$ and $\ell_{sign}$) of steps by $\mathcal{A}$, $V$ outputs its decision bit $b = 1$ meaning that attestation of $S$ was successful, or $b = 0$ otherwise. The result is the output of $V$, i.e., $Exp_{\mathcal{A}} = b$. Thus, a swarm attestation scheme is secure if 
$Pr[b=1|Exp_{\mathcal{A}}(1^{\ell})=b]$ is negligible in $\ell = f(\ell_{N}, \ell_{g}, \ell_{mac}, \ell_{sign})$, where function $f$ is polynomial in $\ell_{N}, \ell_{g}, \ell_{mac}$ and $\ell_{sign}$.
\end{definition}

\begin{definition}
Pseudorandom Function (PRF)

The function $H$ : $K_H \times X \to Y$ can be considered as a secure pseudorandom function
\cite{37}, and $K_H$ is the key space of key $K$ and $X \subseteq \{0, 1\}^{n_1}$ and $Y \subseteq \{0, 1\}^{n_2}$, where $n_1$ and $n_2$ are security parameter and $n_1 > n_2$. The advantage of adversary $\mathcal{A}$ is
\begin{equation}
    \mathrm{Adv}^\mathrm{PRF}_\mathcal{A} = \mathrm{Pr}[\theta = \theta^\prime] - \frac{1}{2}
\end{equation}
where $\theta$ is the output of the truly random function or the pseudorandom function $H$ and $\theta^\prime$ is a guess that $\mathcal{A}$ observes the output. Hence, in any polynomial time, $\mathcal{A}$ in distinguishing $H$ from any given truly random function has a negligible advantage.
\end{definition}

\begin{definition}
Pseudorandom Permutation (PRP)

The function $G$ : $K_G \times Z \to Z^\prime$ can be considered as a secure pseudorandom permutation \cite{38}, and $K_G$ is the key space of key $K$ and $Z \subseteq \{0, 1\}^{n}$, where $n$ is security parameter. The advantage of adversary $\mathcal{A}$ is
\begin{equation}
    \mathrm{Adv}^\mathrm{PRP}_\mathcal{A} = \mathrm{Pr}[\theta = \theta^\prime] - \frac{1}{2}
\end{equation}
where $\theta$ is the output of the truly random permutation or the pseudorandom permutation function $G$ and $\theta^\prime$ is a guess that $\mathcal{A}$ observes the output. Hence, in any polynomial time, $\mathcal{A}$ in distinguishing $G$ from any given truly random permutation has a negligible advantage.
\end{definition}

\begin{definition}
Mutual authentication and authenticated key exchange

An probabilistic polynomial time (PTT) adversary $\mathcal{A}$ and consider two oracles, $\Pi_{A, B}^s$ and $\Pi_{B, A}^t$ where $A$ and $B$ denotes two entities in the communication session $s$ and $t$. The advantage of breaking the mutual authentication $\Pi$ of $\mathcal{A}$ is to break the security of PRP. As a result, the probability of breaking the security of mutual authentication is negligible.
\end{definition}

\begin{definition}
Multi-factor access control

Multi-factor access control uses a multi-factor key method to prevent malicious access by the adversary $\mathcal{A}$. The entity $A$ and the adversary $\mathcal{A}$ each have a key ($pwd_i$ and $k_i$), so $\mathcal{A}$ cannot access the data without $ A$'s authorization. When $A$ accesses its data, it computes a multi-factor and generates the new key for $\mathcal{A}$. $ A$'s key encrypts $ A$'s multi-factor lock. As a result, $\mathcal{A}$ can not know it until $A$ updates the multi-factor lock and responds to the server. 
\end{definition}

\begin{definition}
Against TOCTTOU attack

The adversary $\mathcal{A}$ can read the correct software configuration digest to make attestation successful and perform malicious attacks quickly. Any malicious attacks performed by $\mathcal{A}$ will be detected after the next attestation period. Therefore, it can defend against TOCTTOU attacks.
\end{definition}

\section{Proposed Protocols}
\label{sec:Proposed_Protocol}

We divide the proposed protocol into three phases and detail each entity's interactions in each phase. In addition, notations for our scheme are shown in Table \ref{table:Notations}.

\begin{table}[tb]
\centering
\caption{Notations}
\label{table:Notations}
\begin{tabular}{|c|c|}
\hline
Notation & Description \\
\hline\hline
$\mathrm{PWD}_{tpm}$ & master password for TPM \\\hline
$\mathrm{PWD}_{ss}$ & membership password for cloud IoT server \\\hline
$uid_i$ & unique identity of device $i$ \\\hline
$pwd_i$ & unique TPM password of device $i$ \\\hline
$fd_i$ & firmware / Operating System (OS) digest of device $i$ \\\hline
$AIK_i$ & attestation identity key of TPM \\\hline
$k_i$ & authentication key shared with server and device $i$ \\\hline
$ck_i$ & \makecell{one-time session key shared with \\ device owner and device $i$} \\\hline
$CK_{seed}$ & seed for generating session key \\\hline
$s_i$ & secret for generating session key \\\hline
$\zeta_i$ & encrypted TPM password of device $pwd_i$ \\\hline
$H_i$ & \makecell{keyed hash-based message authentication \\ codes with nonce $n_i$} \\\hline
$g$ & Generator point of ECDH \\\hline
$t$ & time limit for devices joining and attesting \\\hline
$DL, sDL$ & device list and selected device list \\\hline
$VSS$ & valid software status \\\hline
$\mathrm{Enc}_k(x)$ & symmetric encryption with inputs $x$ by key $k$ \\\hline
$\mathrm{Dec}_k(x)$ & symmetric decryption with inputs $x$ by key $k$ \\\hline
$cert(K)$ & \makecell{the certificate of $K$ that can be \\ used to prove the legality of $K$} \\\hline
$ip_i$ & the IP address of device $i$\\
\hline
\end{tabular}
\end{table}

\subsection{Initialization Phase}
The initialization phase, shown in Fig. \ref{fig:Init_Phase}, produces the required parameters and initializes the TPM.  First
%
, we assume the IoT device was initialized offline or is initialized online.  When the device is initialized online, we assume there is a secure channel 
between the device owner and the IoT device. 
The initialization process consists of two steps:

$\ ${\bf Step 1:}
The device owner~(DO) sets a master TPM password $PWD_{tpm}$, generates a one-time session key $ck_i$, a session secret $s_i$, and an ECDH key $x_i$. The TPM uses the session secret to calculate the session key. The IoT device then shares the session key with the DO. Then, the DO uses $PWD_{tpm}$ to derive a unique TPM password $pwd_i$ for the IoT device $i$ using the device's unique identification code~(UID) $uid_i$, i.e., $pwd_i = H(PWD_{tpm}||uid_i)$. Finally, the DO sends $\{pwd_i,ck_i, s_i, x_i\}$ to the IoT device for initialization.

$\ ${\bf Step 2:}
When the IoT device $i$ receives $\{pwd_i,ck_i,s_i,x_i\}$, it uses $pwd_i$ to set a password authorization for the TPM. The TPM generates an attestation identity key $AIK_i$ to ensure the TPM generates a unique secret. Next, the TPM reads the digest of the device's firmware, $fd_i$, and seals the secret $s_i$ in its storage space. Then, it stores the key $x_i$. Finally, the IoT device sends its information, including its IP address $ip_i$ and attestation key certificate $AIK_i$, back to the DO; i.e., the DO is sent $msg_{resp}$, where $msg_{resp}=\{uid_i, ip_i, AIK_i, fd_i, cert(AIK_i), x_i\}$.

\subsection{Registration Phase}
\begin{figure*}[!t]
\centering
\includegraphics[width=\textwidth]{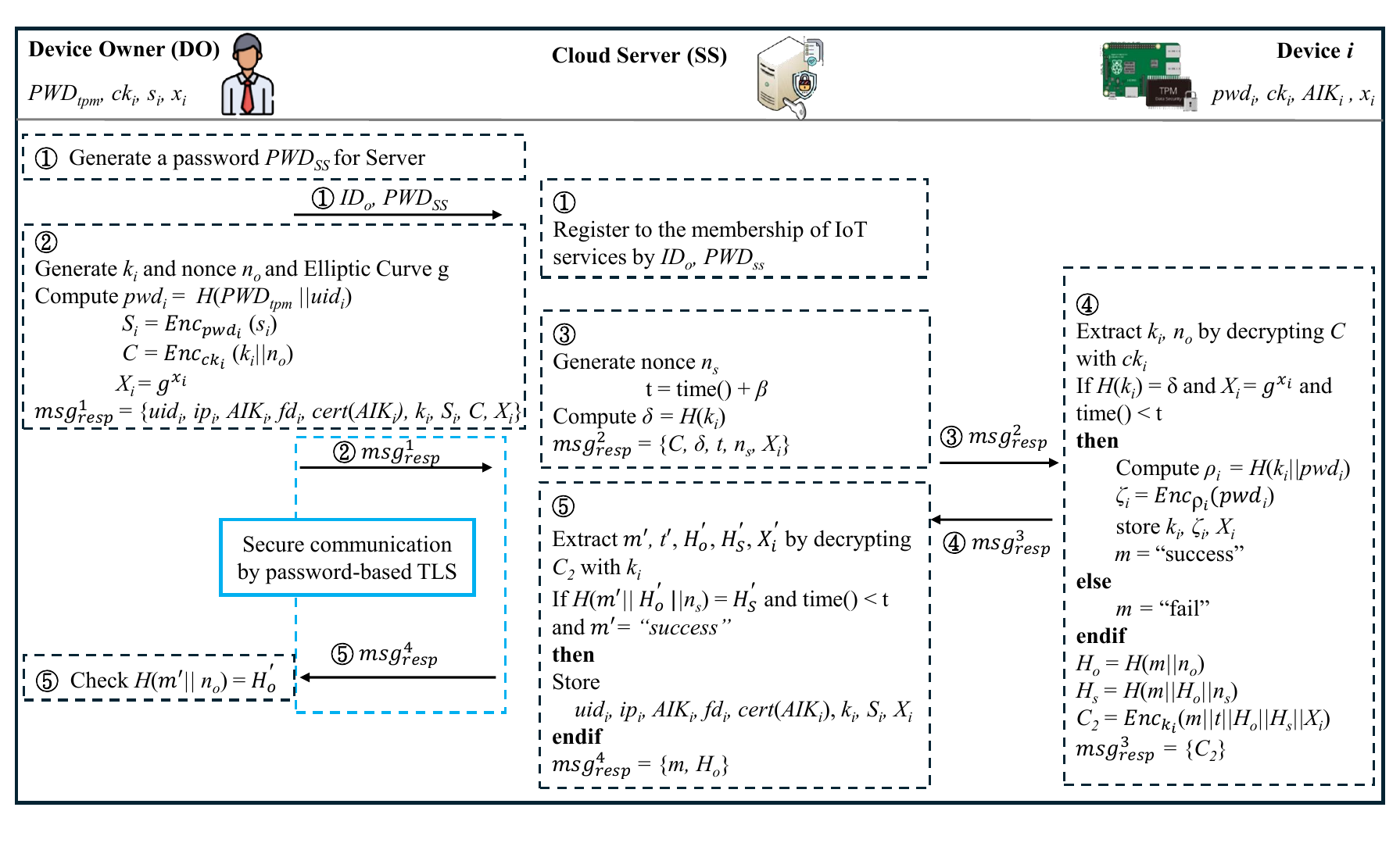}
\caption{Registration phase: In the first step, the device owner generates a password and identity to register at the cloud server. Then, the device owner generates an authentication key, nonce, and session secret and sends them to the cloud server, which generates a nonce, authentication key, and time limit. The device then verifies the device owner's information and sends a hashed, encrypted reply to the cloud server. The cloud server decrypts and verifies the nonce and time limit. If they are consistent, the device is registered. After which, the device owner is notified.}
\label{Fig:Reg_Phase}
\end{figure*}
Figure 7 shows that the DO registers an IoT device on the cloud IoT service. The DO can confirm whether the device is registered as follows:

$\ ${\bf Step 1:}
The DO first generates a password, $PWD_{ss}$, and an identity, $ID_o$, to register as a member of the cloud IoT service.

$\ ${\bf Step 2:}
The DO then generates a cloud IoT service authentication key, $k_i$, and a nonce, $n_o$. The former allows the device to authenticate to an authorized cloud IoT service server, and the nonce ensures response freshness and prevents replay attacks by using previous attestation results. Afterward, the DO encrypts the session secret, $s_i$, shared with the IoT device using the unique TPM password $pwd_i$.  Then the DO uses the one-time key $ck_i$ to encrypt $k_i$ and $n_o$, thus preventing the cloud IoT service server from accessing the IoT service. Next, the DO uses key $x_i$ to generate value $X_i$. Finally, the device information ($uid_i$, $ip_i$, $fd_i$), TPM attestation identity key $AIK_i$, authentication key certificate $cert(AIK_i)$, authentication key $k_i$, the encrypted session key $S_i$, the encrypted IoT service authentication key $C$, and value $X_i$ are sent to the cloud IoT service server.

$\ ${\bf Step 3:}
When the cloud server receives DO's messages, it generates a nonce $n_s$ and a time limitation period $t$. Afterward, it hashes the authentication key $\delta = H(k_i)$. Finally, it sends $C, \delta, t, n_s, X_i$ to the device via the IP address in the device information.

$\ ${\bf Step 4:}
When the IoT device receives the cloud server's messages, it first decrypts $C$ via the one-time key $ck_i$ to obtain the authentication key $k^{'}_i$ and the nonce $n_o$.  Then it hashes the authentication key $k^{'}_i$ to compare whether it is the same as $\delta$, and it also compares $X_i=g^{x_i}$. 
If all are the same, then the IoT device calculates $\rho_i = H(k_i||pwd_i)$ and uses $\rho_i$ to encrypt the TPM password $pwd_i$ as $\zeta_i$. Finally, it generates the message "success" and hashes it with the device owner's name $n_o$. 
The result $H_o$ is hashed with the server's nonce $n_s$ and time limitation period $t$. Finally, it uses the authentication key $k_i$ 
to encrypt the message and hashed value.  Then the IoT device sends $C_2$ back to the server.

$\ ${\bf Step 5:}
When the server receives the response, it decrypts $C_2$ and $k_i$ to extract $m^\prime, t^\prime, H^\prime_o, H^\prime_s, X^\prime_i$. Next, the server compares whether the received response $H_s^{'}$ is the same as using the nonce $n_s$ to hash $H(m^\prime||H_o^{'}||n_s)$ and whether time $t$ has expired. If both are consistent, the server registers the device information and sends $m$ and $H_o$ to the DO. The DO verifies whether the device registration is successful.

\subsection{Attestation Phase}
In this phase, the DO logs in to the cloud IoT service to select devices for attestation. As shown in Fig. \ref{fig:Attestation_phase}, the DO uses the multi-factor key algorithm to resist possible malicious recording behavior on the server and authenticates the session key exchange during attestation as described as follows:

\begin{figure*}[!t]
\centering
\includegraphics[width=\textwidth]{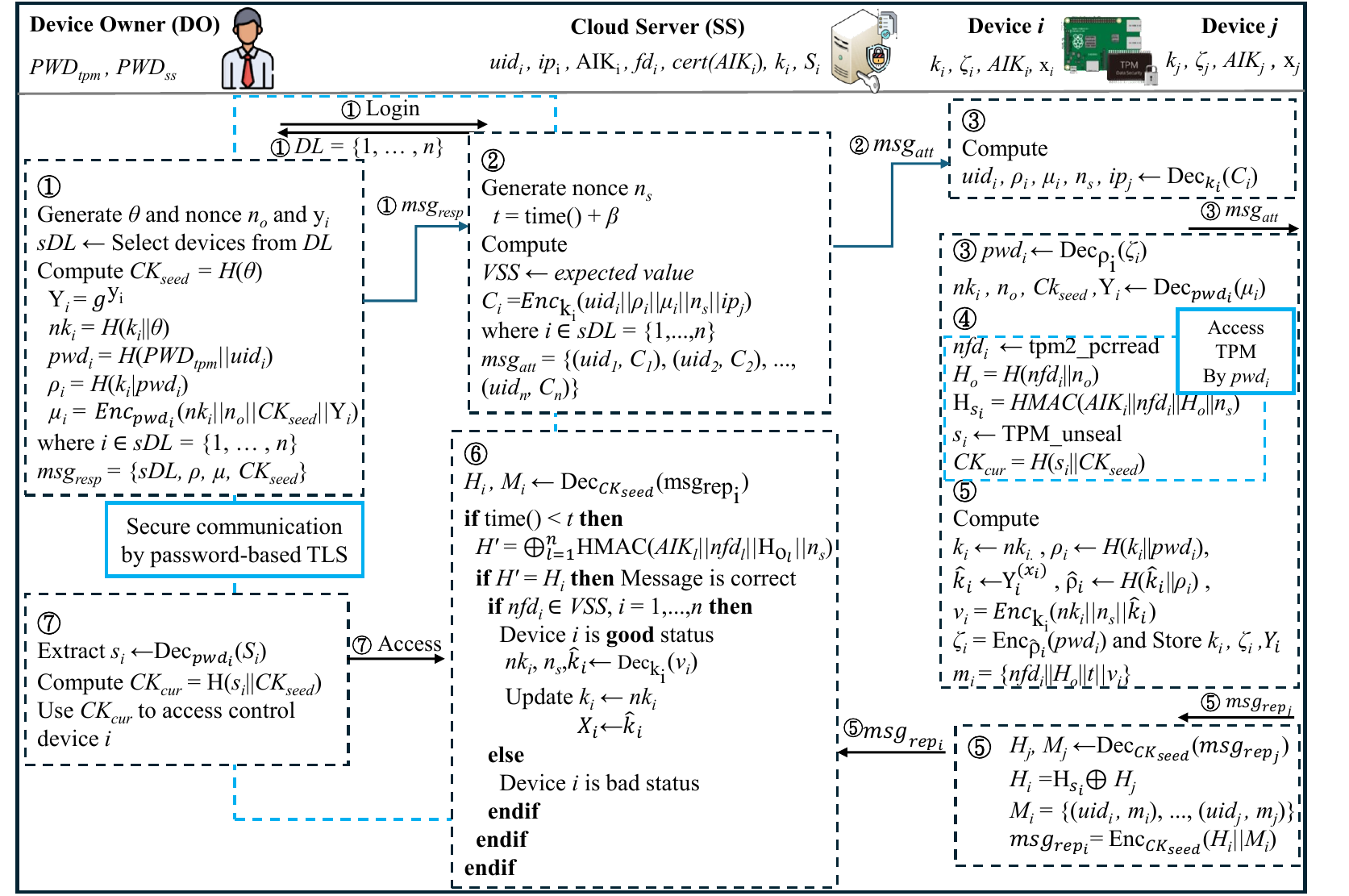}
\caption{Attestation phase: The device owner logs into the cloud server and generates an attestation request for each selected device.  The cloud server then generates a device-specific attestation request for each selected device. The recipient device decrypts the attestation request and sends it to each connected node. Each device then obtains its firmware digest and sends its attestation response to its parent node. The parent node sends its aggregated attestation responses to the cloud server. The cloud server then verifies the responses, checks the software status, and updates each device's status on the server. The device owner can verify the report's authenticity and determine which of the devices failed to attest correctly.}
\label{fig:Attestation_phase}
\end{figure*}

$\ ${\bf Step 1:}
The DO logs in to the cloud server, selects the devices to be attested, and generates a secure random number $\theta$, a nonce $n_o$, and a new ECDH key $y_i$. Next, the server uses $\theta$ to calculate the seed for the session key $CK_{seed} = H(\theta)$ and the new authentication key $nk_i = H(k_i||\theta)$. Afterward, the server computes $Y_i=g^{y_i}$. Then, based on the server-provided IoT device information, the DO computes $\rho$ and $\mu$ for each device. Finally, the DO sends the attestation verification device to the cloud server so it can generate attestation requests.

$\ ${\bf Step 2:}
After receiving the DO's messages, the server generates a nonce $n_s$ and a time limit $t$ per $time()$ and $\beta$. Then the server encrypts each message individually with device-specific authentication keys, computes the devices' valid software status $VSS$, and generates a tree network topology. The server sends the encrypted request to the selected devices through the network topology. 

$\ ${\bf Step 3:}
When the first device $i$ receives the request, it uses the authentication key $k_i$ to decrypt the ciphertext $C_i$ and forwards the remaining requests to the subsequent nodes using their IP addresses, within the context of $C_i$. Next, the IoT device decrypts $\zeta_i$ to get the unique TPM password $pwd_i$ and uses it to decrypt $\mu_i$ to obtain a new authentication key $nk_i$,  which is then assigned to $k_i$, and generates a nonce $n_o$ and a value $Y_i$. 

$\ ${\bf Step 4:}
The IoT device uses the unique TPM password $pwd_i$ to enable the TPM and obtain the current firmware digest $nfd_i$. In the TPM's secure computing space, it computes $H_o$ and uses the attestation identity key $AIK_i$ to perform keyed-hash MAC operations on the firmware digest and nonce. Then, the TPM unseals the session secret $s_i$ to compute a new session key $CK_{cur} = H(s_i||CK_{seed})$.

$\ ${\bf Step 5:}
The IoT device uses the authentication key $k_i$ to encrypt the new authentication key $nk_i$ and nonce $n_s$ and then uses $nk_i, x_i, Y_i$ to update the stored parameters $k_i$ and $\zeta_i$. Afterward, the IoT device waits for the child node's attestation response for a limited time; if it receives no response, it encrypts the MAC tag and message with $CK_{seed}$ (without the leaf node response) and sends them to the upper branch. If the leaf node responds, the IoT device decrypts the response using $CK_{seed}$, aggregates the leaf node's MAC tag with the other MAC tags to form a new tag, and concatenates the messages. Finally, the device encrypts the MAC tag and message using $CK_{seed}$ to respond to an upper branch or the cloud server.

$\ ${\bf Step 6:}
When the cloud server receives the message, it decrypts the response via $CK_{seed}$ and confirms whether it is within the permitted time limit $t$. If within the time limit, the cloud server will perform a MAC 
verification and use the information in its message to form a new tag $H^\prime$ to verify that the received tag $H_i$ is correct and complete. If it is correct, the cloud server will check the firmware digest of each selected device to confirm whether it is in a valid software status $VSS$. If the software status is confirmed, the cloud server sets the corresponding IoT device status to "good," replaces its authentication key $k_i$ with $nk_i$, and replaces $X_i$ with $\hat{k}_i$. Otherwise, the server sets the device status to "bad." Finally, regardless of the device's status, the server updates the IoT device's firmware digest and hashed digest.

$\ ${\bf Step 7:}
The DO accesses the cloud server and verifies whether the server is forging the device's status. The DO decrypts the encrypted secret seed $S_i$ stored on the cloud server to get $s_i$ and calculates a new session key $CK_{cur} = H(s_i||CK_{seed})$ to access each IoT device.

If the aggregated attestation report fails to pass verification, it is difficult to distinguish which attestation report is invalid. Therefore, we can use a mechanism like recording log files that alleviates the impact of failed verification, such as batch verification, to establish device checkpoints. 
\section{Security Analysis and Comparison}
\label{sec:Sec_Analysis_Compare}
In this section, we analyze the security requirements our protocol meets and compare key security features with prior work. Finally, we evaluate the feasibility of our scheme through implementation and simulation.

\subsection{Security Analysis}
In this section, we will prove the security of the proposed protocol against 
passive malicious administrators as follows.
\newtheorem{theorem}{Theorem}
\begin{theorem}
Our scheme achieves secure swarm attestation if 1) the underlying MAC scheme is selective forgery resistant, 
2) The MAC scheme's hash function $H$ is pseudorandom, and 3) symmetric encryption $E_s$ is a pseudorandom permutation.
\end{theorem}
\begin{proof}
The verifier $\mathcal{V}$ receives only $msg_{rep}$, encrypted with a session key sent by the device owner. The plaintext content of $msg_{rep}$ is $(H||M)$, where $H$ is the aggregated MAC and $M$ is the remaining parameters such as $nfd$, the reference software configuration, and the nonce $N$ sent by $\mathcal{V}$ and the timestamp $t$. We suppose that the adversary $\mathcal{A}$: 1) can tamper with the software of any device $D \in S$ and 2) can eavesdrop and modify the messages between $D \in S$ and $\mathcal{V}$.
\end{proof}

In the first case, $\mathcal{A}$ can modify the device's software. In our scheme, since our devices are equipped with a secure hardware module TPM, to measure software configuration, 
$\mathcal{A}$ can not modify with code performing integrity measurements $nfd$ on the device and attestation report. If $\mathcal{A}$ modifies the measurements $nfd^\prime$, it differs from $nfd$ in the MAC $H(nfd)$ performed by the TPM. Thus $\mathcal{A}$ must forge MAC $H$ to make $\mathcal{V}$ accept. According to {\it Definition 2}, if the hash function in the MAC scheme is random, the probability $\ell_{mac}$ that $\mathcal{A}$ can successfully forge MAC $H$ is negligible.

For the second case, $\mathcal{A}$ can eavesdrop and modify the response message. The messages transmitted by both entities are encrypted by different session keys in each attestation, so $\mathcal{A}$ can not decrypt the messages if $\mathcal{A}$ does not know the session key. By {\it Definition 3}, if symmetric encryption is a random permutation, the probability that $\mathcal{A}$ can successfully modify the message is negligible.

Therefore, we can conclude that the probability of $\mathcal{A}$ making $\mathcal{V}$ accept in the case where $\mathcal{A}$ modified the measurements of the software configuration of the device in $\mathcal{S}$ is negligible.

\begin{theorem}
Our scheme achieves security against passive malicious administrators if the hash function $H$ is a pseudorandom function with PRF security and symmetric encryption $E_s$ is a pseudorandom permutation with PRP security.
\end{theorem}
\begin{proof}
Suppose that $\mathcal{A}$ is an adversary that breaks the security against passive malicious administrators in our proposed protocols in probabilistic polynomial time. Thus, to obtain the key, $\mathcal{A}$ may use the following strategies: 1) using $\rho$ sent by the device owner to obtain $pwd_i$; 2) extracting $s$ stored on the server to compute a session key.
\end{proof}

In the first case, the server receives $\rho$ during the attestation phase, and $\mathcal{A}$ can record the current value and try to obtain the original value. $\rho$ is an output that hashes a selected value $k_i|pwd_i$. By {\it Definition 2}, if the hash function is random, $\mathcal{A}$ has no advantage in breaking security against passive malicious administrators.

In the second case, $S$ is stored on the server because the device owner registered the device with the server, and $\mathcal{A}$ can extract $s$. $S$ is a ciphertext under symmetric encryption, Advanced Encryption Standard (AES-256). Thus, by {\it Definition 3}, if symmetric encryption is a random permutation, $\mathcal{A}$ has no advantage in breaking the security against passive malicious administrators.

Therefore, the probability of breaking the security against passive malicious administrators is negligible in the above cases, and our proposed protocol is secure against passive malicious administrators with the IND-CCA 
security of the underlying pseudorandom function and the pseudorandom permutation.\\

\subsubsection{Mutual authentication}
We assume the RA protocol $\Pi$ is a secure mutual authentication protocol for the server, and all devices under the pseudorandom permutation exhibit PRP security.

\begin{proof}
Suppose that $\epsilon$ is the probability of breaking the mutual authentication of $\Pi$ by a probabilistic polynomial-time (PPT) adversary $\mathcal{A}$ with the simulator $S$, who simulates $\Pi$ and plays the game to break the PRP security. $S$ simulates the server or the device with a given pair of $\mathrm{PRP}(\Omega, \Omega^{-1})$ by a challenger $\psi$ to interact with $\mathcal{A}$. $S$ simulates $\Pi$ by the following oracles to capture the ability of $\mathcal{A}$, where $A$ denotes the server, $B$ denotes the device in $\Pi$, and $s$ and $t$ denote the communication session.
$\bullet$
$\mathrm{Execute}(\Pi^s_{A,B}, \Pi^t_{B,A})$: The oracle models the execution of $\Pi$ and outputs the executed script.\\
$\bullet$
$\mathrm{Send}(\Pi^s_{A, B}, m)$: The oracle models an active attacker, who sends a message $m$ to
$\Pi^s_{A,B}$. If $m = \{C\}$, $\Pi^s_{A,B}$ will decrypt $C$ by $\Omega^{-1}$ and if the content of plaintext is $(uid, \rho, \mu, n_s, ip)$, it means the message is correct. Then, the oracle will output accept to authenticate this session or reject the session otherwise.\\
$\bullet$
$\mathrm{Send}(\Pi^t_{B,A}, m)$: The oracle is similar to the above. If $m = \{C_2\}$, $\Pi^t_{B,A}$ will also check whether the plaintext is $(H, uid, msg)$. If so, the oracle will also authenticate the session and perform verification actions.
\end{proof}
{\bf Challenge: }Before the challenge phase, $S$ can use the polynomial number of messages as input to query $\Omega$ and $\Omega^{-1}$. It also receives other corresponding outputs for training. $S$ sends
a chosen message $m = uid, \rho, \mu, n_s, ip$ or $m = (H, uid, msg)$, which as above message in the simulated oracle with $\mathcal{A}$, to $\psi$. Then, $\psi$ will randomly choose a bit $b \in \{0,1\}$. If $b$ = 0, then $\psi$ uses $\Omega$ to encrypt $m$; If $b$ = 1, then $\psi$ uses $\omega$ to encrypt $m$. Finally, it outputs the result to $S$.

{\bf Guess: }
If $\mathcal{A}$ sends a legal message $C$ or $C_2$ as the correct interaction between the server and the device, $S$ will output $b^\prime = b = 0$; otherwise, $S$ will output $b^\prime \in \{0, 1\}$ randomly. When $b$ = 0, the above experiment is a real experiment, and $\mathcal{A}$ has additional advantages that break the mutual authentication of $\psi$; otherwise, when $b$ = 1, the experiment is a random experiment, and $\mathcal{A}$ has no advantage in breaking the mutual authentication of $\psi$. Therefore, $S$ can only use $\epsilon$, 
the advantage of breaking the mutual authentication $\Pi$ of $\mathcal{A}$, to break the security of PRP. Hence, we have that $\tau \ge Pr[b^\prime = b] - \frac{1}{2} = \epsilon + \frac{(1 - \epsilon)}{2} - \frac{1}{2} = \frac{\epsilon}{2}$. Because $\tau$ is negligible by {\it Definition 3}, $\epsilon$ is also negligible by the above equation. Thus, the probability of breaking the security for mutual authentication is negligible, and our protocols are secure.

\begin{table*}[tb]
\centering
\caption{Comparisons on security properties}
\label{table:Compare_Security_Prop}

\begin{tabular}{|c|c|c|c|c|c|c|c|c|}
\hline\hline
&
\cite{12}, \cite{44},\cite{45}
&
\cite{14}
&
\cite{40},\cite{frolikov2025toctou}
&
\cite{41}
&
\makecell[l]{\cite{07}, \cite{08}, \cite{10}, \\ \cite{42}, \cite{43}}
&
\cite{11},\cite{shen2026ltraa}
&
\cite{sekhavati2025resisting}
&
Ours
\\
\hline\hline

Integrity of Attestation Data
& \Checkmark & \Checkmark & \Checkmark & \Checkmark & \Checkmark & \Checkmark & \Checkmark & \Checkmark \\

Selectable Remote Attestation
& \Checkmark & \XSolid & \XSolid & \Checkmark & \XSolid & \Checkmark & \XSolid & \Checkmark \\

Session Key Exchange during RA
& \XSolid & \XSolid & \XSolid & \XSolid & \XSolid & \Checkmark & \Checkmark & \Checkmark \\

Against TOCTTOU Attack
& \XSolid & \Checkmark & \Checkmark & \XSolid & \XSolid & \XSolid & \XSolid & \Checkmark \\

Against Unauthorized Access
& \XSolid & \Checkmark & \XSolid & \Checkmark & \XSolid & \XSolid & \Checkmark & \Checkmark \\

Trust of Verifier
& $\blacksquare$ & $\blacksquare$ & $\blacksquare$ & $\blacksquare$ & $\blacksquare$ & $\square$ & $\square$ & $\square$ \\

Type of Verifier
& $\bigcirc$ & $\bigcirc$ & $\bigcirc$ & $\bigcirc$ & $\bigcirc$ & $\triangle$ & $\triangle$ & $\triangle$ \\

\hline\hline
\end{tabular}

\smallskip

\begin{tabular}{|c|}
\hline
$\blacksquare$: Fully trust, $\square$: Semi-trust, $\bigcirc$: Local, $\triangle$: Cloud \\
\hline
\end{tabular}
\end{table*}

\subsubsection{Authenticated key exchange}
In our protocol, the device owner and the device can exchange a key during the remote attestation process to compute a session key $CK_{cur}$. First, when the device owner makes an attestation request, the device generates a session seed $CK_{seed}$ and encrypts it as $\mu$. It then sends it to the device with other messages. This means others cannot tamper with the session seed. During the attestation protocol, the device obtains the session seed by decrypting the ciphertext and calculating a new session key with the secret seed unsealed from the TPM: $CK_{cur} = H(s_i||CK_{seed})$.

As described above, our protocols use symmetric encryption to protect the key material from $\mathcal{A}$. Thus, $\mathcal{A}$ has no advantage in breaking the security of authenticated key exchange under the PRP security by {\it Definition 3}.

\subsubsection{Multi-factor access control}
Our system uses this multi-factor key method to prevent malicious server access. The IoT device has a multi-factor lock that requires multiple keys to decrypt and execute successfully. However, the device owner and the server each have a key ($pwd_i$ and $k_i$), so the server cannot access the device without the device owner's authorization. When the device owner wants to access the device through the server, although the device owner will compute a multi-factor key $\rho$, the device owner will also generate a new authentication key $nk_i$, which will replace the server's authentication key $k_i$ to update the multi-factor lock of the device and encrypt it with TPM password $pwd_i$ so that the server can not know it until the device updates the multi-factor lock and responds to the server.\\

According to {\bf Theorem 2} and the above analysis, multi-factor access control also meets {\it Definition 2} and {\it Definition 3}. Hence, {\bf Theorem 2} covers the security of multi-factor access control.

\subsubsection{Attestation data integrity}
In our scheme, the TPM generates the attestation data using the devowner's nonce: first, $H_o = H(nfd||n_o)$. Then, the attestation data is computed as a keyed-hash MAC using $H_o$, the server nonce, and the authentication key $ H_S = H(AIK||nfd||H_o||n_s)$. In this way, the data integrity is doubly guaranteed. In other words, we assume the TPM is a fully trusted security device. Therefore, the attestation data is fully trusted. Next, it computes a MAC using the device owner's nonce, so the device owner can directly verify the integrity of the attestation data and ensure it has not been tampered with, since only the TPM knows the device owner's nonce. Furthermore, the TPM uses an AIK authenticated by the CA to compute a keyed-hash MAC over the attestation data, $H_o$, and the server nonce, $H_s = H(AIK||nfd||H_o||n_s)$. Consequently, when the attestation data is sent to the server, the server can verify its integrity. However, it cannot tamper with the data because $H_o$ still verifies the integrity of the attestation data for the device owner.\\
By {\bf Theorem 1} and the above analysis, the attestation data also meets {\it Definition 2}. Thus, the security of the integrity of attestation data is within {\bf Theorem 1}.

\subsubsection{Freshness}
In our protocols, each message carries the server and device-owner nonces $n_s$ and $n_o$, and each attestation request has a limited time $t$. Therefore, we can check whether the nonce matches to ensure the response is fresh. Furthermore, when the device owner requests attestation, they will generate a new authentication key $nk$ and a new session seed $ck_{seed}$. Therefore, after running the protocol, the attested device updates its authentication key and generates a new session key. The server also updates the authentication key for the corresponding device. $\mathcal{A}$ offers no advantage in breaking freshness because each timestamp (nonce and update key) uses a random number or hash function and meets {\it Definition 2}.
\subsubsection{Against TOCTTOU attack}
This TOCTTOU attack uses legal software configuration to let the device read the correct software configuration digest, making the attestation successful and enabling malicious attacks most of the time. According to the research \cite{14}, they use TPM to defend against TOCTTOU attacks, and our scheme also uses TPM to measure software configuration. This means that when an attack changes the device's configuration, the TPM's PCR value reflects it. Because the PCR value changes, the TPM cannot unseal the secret seed $s_i$, and we can then tell that the device has been compromised. Therefore, it satisfies {\it Definition 6}. Furthermore, if the TPM successfully unseals the secret seed $s_i$, the current device is not attacked, so the new session key is safe.\\
Additionally, based on {\it Definition 6} and some research \cite{rfc9334, banks2021}, the attestation interval should account for the threat model, required security, and the computational resources of attestation devices. However, these studies do not provide a practical standard for determining the attestation interval. 
Keylime \cite{Keylime}, a TPM-based remote attestation framework, specifies a fixed attestation interval. Although Keylime lets users configure the attestation interval, its default is 2 seconds. However, completing the entire attestation process within 2 seconds is challenging in large-scale IoT environments, especially with communication delays.
Based on the above considerations, we set the attestation interval from 30 minutes to one hour. Besides, the device needs to check its PCR value periodically. The checking interval can be set from 2 to 5 seconds. If the device detects a change in the PCR value, it must respond to the verifier. The response message should include at least $AIK$ and a nonce, with a hash operation for integrity verification.
\subsection{Comparisons on Security Features}
In this section, we compare the security properties with these related works: SEDA \cite{07}, WISE \cite{12}, and MTRA \cite{14}, as shown in Table \ref{table:Compare_Security_Prop}.

First, because these three schemes relate to the security properties of our schemes, we compare them. SEDA is the initial swarm RA protocol, WISE is the intelligent swarm RA protocol, and MTRA uses TPM for RA.

For the integrity of attestation data, every research related to RA is included because it is necessary to ensure the correctness of the attestation report. Both our scheme and WISE can select devices for selectable remote attestation. WISE classifies devices and can flexibly select the devices wanted for RA. For session key exchange during RA, we design a method that lets the device receive materials from the device owner without the server knowing. Against TOCTTOU attacks, our scheme and MTRA use the TPM for RA. TPM can effectively prevent such attacks. To prevent unauthorized access, MTRA uses the hash-chain method, so the verifier must know its value to access and attest the device. Regarding verifier trust, the other three works fully trust the verifier, or the verifier is the device owner. For the other works, the verifier is always near the device.
\subsection{Implementation cost of TPM}
In this section, we discuss the feasibility of deploying TPMs on all IoT devices. While staffing and implementation costs are difficult to estimate, we mainly focus on component cost. For component cost, the unit price for a bulk package of 4000 units of the Infineon SLS32AIA, a generic TPM module, ranges from 0.20 to 0.80 US dollars per unit. This price is the upper bound, as large companies typically pay less because vendors are willing to discount beyond what the aforementioned distributors provide. If an IoT device can deploy a trusted virtual TPM, then it can be implemented on many devices at an estimated end-user price of 1.21 US dollars per IoT device, which is practical for large companies and end users.
\section{Evaluation}
\label{sec:Implement_Perf_Eval}
In this section, we implement a prototype system for our proposed scheme to evaluate real-world performance and assess feasibility in a simulated environment with many IoT devices.
\subsection{Implementation}
\begin{figure}[!t]
\centering
\includegraphics[width=1\columnwidth]{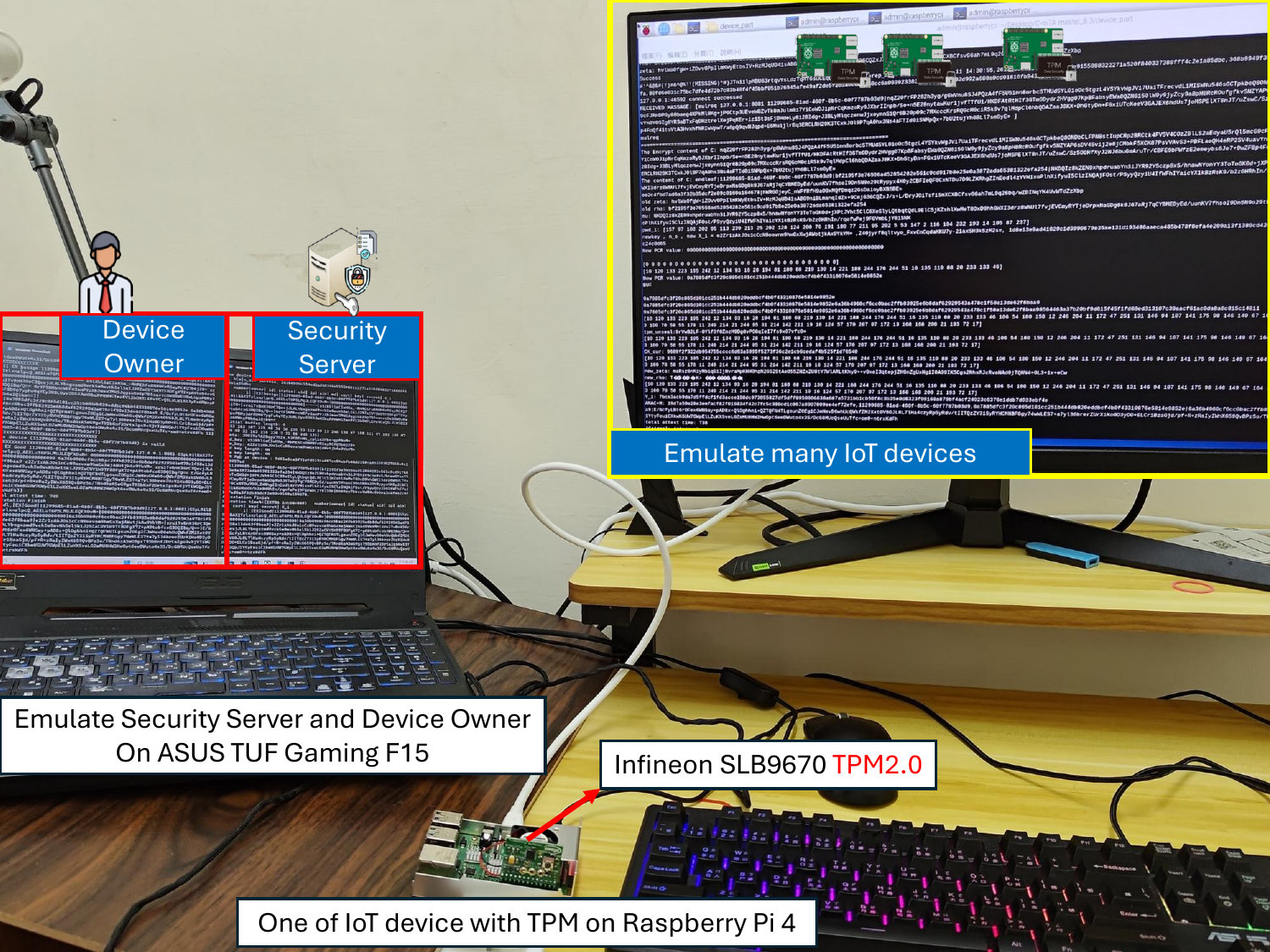}
\caption{Prototype system deployment. The device owner initializes the IoT devices directly and then registers them via the security server. Afterward, the device owner manages all IoT devices via the security server.}
\label{fig:Deployment}
\end{figure}

As shown in Fig. \ref{fig:Deployment}, we implement our proposed scheme to measure the computation costs on a Raspberry Pi 4 and a laptop. We use Golang (version go1.15.5, darwin/amd64) to implement the client (device owner) and the server on an ASUS TUF Gaming F15 laptop with an Intel Core i5 CPU (2.5GHz to 4.5GHz) and 16 GB of memory. The OS is Windows 11, version 25H2. The Raspberry Pi 4 runs Debian GNU/Linux 12. We also use the Golang programming language (version go1.13.1 on Linux) to deploy the IoT application on a Raspberry Pi 4 Model B with an ARM Cortex-A72 (v8) 64-bit CPU, a 1.5GHz clock rate, and 4 GB of memory. In
addition, the hardware module TPM we use is the Infineon SLB9670 TPM2.0.

\subsection{Computation Costs}
\begin{table}[b]
\centering
\caption{Computation costs on Raspberry Pi 4}
\label{table:Compute_RPi_Costs}
\begin{tabular}{|c|c|}
\hline
Initialization & 1772.57 ms \\ \hline
Registration & $2.99529 \times 10^{-1}$ms \\ \hline
Attestation$^*$ & 528.536 ms \\ \hline
Aggregate HMAC & $5.6887 \times 10^{-2}ms$ \\ \hline
Symmetric encryption & $8.7703 \times 10^{-2}ms$ \\ \hline
ECDH encryption & $1.56 \times 10^{-1}ms$ \\
\hline
\multicolumn{2}{l}{$^*$The aggregate HMAC operation is not included here} \\
\end{tabular}
\end{table}

Table \ref{table:Compute_RPi_Costs} shows that we implemented the IoT device in the protocol with a Raspberry Pi 4. Our protocols use cryptographic operations such as symmetric encryption (AES-256) and hash (SHA-256) operations to reduce the computational burden. Except for initialization and attestation, which take a long time due to TPM operations, the rest takes less than a millisecond. The device owner and server usually have unlimited computing power, and the RA protocol mainly considers the attestation efficiency of low-power devices, so we do not consider them here.

As shown in Table \ref{table:Compare_Computation_Costs_Attest}, we compare total computation costs for attestation with the other related research, SEDA \cite{07}, WISE \cite{12}, TRAIN \cite{frolikov2025toctou}, S. Sekhavati et al. \cite{sekhavati2025resisting}, and MTRA \cite{14}, and show the notations for the computation times of all operations. In our scheme, there are selected device numbers of" N " computations executed on devices with unlimited computing resources, such as the device owner and server. Therefore, this does not affect the runtime of the attestation of IoT devices with low computational power. These computations prevent passive malicious administrators from accessing the secret.

\subsection{Scalability Test}
Because we are in an environment with many IoT devices, we must evaluate whether our scheme is scalable and applicable to the real world. Therefore, we simulate 100000 devices running our protocol across different tree network topologies with a Wi-Fi communication protocol and use measured Raspberry Pi 4 delays as the execution time for each device, with nodes in each layer attested in parallel.

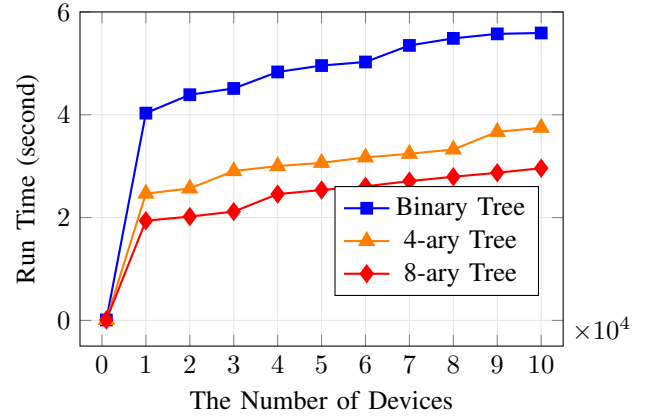
\begin{figure}[!ht]
  \centering
  \begin{tikzpicture}
   
   \begin{axis}[
    width=0.9\linewidth,
    height=6cm,
    xlabel={The Number of Devices},
    ylabel={Run Time (second)},    
    xmin=-0.5,
    xmax=10.5,
    ymin=-0.5,
    ymax=6,
                ylabel style={
                    yshift=0pt
                },
    xtick={0,1,2,3,4,5,6,7,8,9,10},
    ytick={0,2,4,6,8},
    grid=both,
    clip=false,
    grid style={gray!20},
    legend style={
     at={(0.95,0.15)},
     anchor=south east,
     draw=black
    }
    ]

    \node at (axis description cs:1.08 , 0.05) {$\times$$10^4$};
    
    \addplot[
    color=blue,
                mark=square*,
    thick,
    ]
    coordinates {
     (0.1,0.0083401)
     (1,4.03182461)
     (2,4.3898558)
     (3,4.5118558)
     (4,4.8328869)
     (5,4.9568869)
     (6,5.0268869)
     (7,5.3489181)
     (8,5.4859181)
     (9,5.5739181)
     (10,5.5909181)
    };
    
    \addplot[
    color=orange,
                mark=triangle*,
                ,mark size=3pt,
    thick,
    ]
    coordinates {
     (0.1,0.0083401)
     (1,2.46566675)
     (2,2.56666675)
     (3,2.90470282)
     (4,3.00170282)
     (5,3.06470282)
     (6,3.17070282)
     (7,3.24070282)
     (8,3.32470282)
     (9,3.66873889)
     (10,3.74673889)
    };
    
    \addplot[
    color=red,
                mark=diamond*,
                mark size=3pt,
    thick
    ]
    coordinates {
     (0.1,0.0083401)
     (1,1.93863365)
     (2,2.01963365)
     (3,2.11663365)
     (4,2.45667948)
     (5,2.53667948)
     (6,2.60767948)
     (7,2.70967948)
     (8,2.79367948)
     (9,2.87167948)
     (10,2.96067948)
    };
    
    \legend{
     Binary Tree,
     4-ary Tree,
     8-ary Tree
    }
    
   \end{axis}
   
  \end{tikzpicture}
  
  \caption{Run Time Comparison for Different Tree Structures}
  \label{fig:Run-time}
  
\end{figure}

The execution attestation time of 100,000 devices in binary-tree topology is approximately 3.201 seconds. The execution attestation time in the 4-ary tree topology is approximately 2.0723 seconds, and the execution attestation time in the 8-ary tree topology is approximately 1.6734 seconds, as shown in Fig. \ref{fig:Run-time}.

\begin{table*}[thb]
\centering
\caption{Comparisons of computation costs for each attestation}
\label{table:Compare_Computation_Costs_Attest}
\begin{tabular}{|c|c|c|c|}
\hline
Scheme & \multicolumn{3}{|c|}{Computation costs} \\
\hline\hline
 & Owner & Verifier & Device \\
\hline
SEDA \cite{07} & 0 & $T_{versig}$ & \makecell{$n \cdot T_{rng} + T_{sig} +$ \\ $(2n+2)T_{mac} + (3n+1)T_{cp}$} \\
\hline
WISE \cite{12} & 0 & \makecell{$T_{clu}+ (N+1)T_{mac}$} & \makecell{$T_{clu_{cp}} + T_{mac}$ \\ $+ 4T_{cp} + n \cdot T_{or}$} \\
\hline
MTRA \cite{14} & 0 & $2n \cdot T_{se}$ & \makecell{$2T_{se} + T_h + T_{cp}$ \\ $+ 2T_{seal} + T_{unseal} + T_{read}$} \\
\hline
TRAIN\cite{frolikov2025toctou} & 0 & \makecell{$n \cdot T_h + n \cdot T_{mac}$} & \makecell{$T_h + T_{mac}$}\\
\hline
S. Sekhavati et al.\cite{sekhavati2025resisting} & \makecell{$2 \cdot T_h + 1 \cdot T_{xor}$} & \makecell{$ (5N+2)T_{xor}+(8N+3)T_h$ \\$+ N \cdot T_{se}$} & \makecell{$7T_h+6T_{xor}+T_{se}$}\\
\hline
Ours & \makecell{$T_{tpmrng}+T_{ecdh} $ \\ $+ (3N+1)T_h+ N \cdot T_{se}$} & \makecell{$(3n)T_{se}+ n \cdot T_{xor}$} & \makecell{$4T_{se} + 3T_h + T_{mac}+T_{ecdh}$ \\ $ + T_{unseal} +  T_{read}$} \\
\hline
\multicolumn{4}{|l|}{$n$: the number of devices on child nodes in the network topology} \\
\multicolumn{4}{|l|}{$N$: the number of selected devices on this attestation} \\
\multicolumn{4}{|l|}{$T_{clu}$: the computation time of generating the clusters for selecting devices} \\
\multicolumn{4}{|l|}{$T_{clu_{cp}}$: the computation time for comparing whether the device is in the cluster} \\
\multicolumn{4}{|l|}{$T_{rng}$: the computation time of generating random number / nonce} \\
\multicolumn{4}{|l|}{$T_{tpmrng}$: the computation time of generating a random number/nonce by TPM} \\
\multicolumn{4}{|l|}{$T_{sig}$: the computation time of computing a signature (RSA 1024 Signature)} \\
\multicolumn{4}{|l|}{$T_{versig}$: the computation time of verifying the signature (RSA 1024 Verification)} \\
\multicolumn{4}{|l|}{$T_{se}$: the computation time of symmetric encryption / decryption (AES-256)} \\
\multicolumn{4}{|l|}{$T_h$: the computation time of one-way hash function (SHA-256)} \\
\multicolumn{4}{|l|}{$T_{mac}$: the computation time of generating MAC (keyed-hash based MAC)} \\
\multicolumn{4}{|l|}{$T_{or}$: the computation time of OR operator} \\
\multicolumn{4}{|l|}{$T_{xor}$: the computation time of XOR operator} \\
\multicolumn{4}{|l|}{$T_{cp}$: the computation time of comparison function} \\
\multicolumn{4}{|l|}{$T_{seal}$: the computation time of TPM seal function} \\
\multicolumn{4}{|l|}{$T_{unseal}$: the computation time of TPM unseal function} \\
\multicolumn{4}{|l|}{$T_{read}$: the computation time of TPM read software configuration function} \\
\multicolumn{4}{|l|}{$T_{ecdh}$: the computation time of ecdh calculation} \\
\hline
\multicolumn{4}{|l|}{The computation costs of operations on the owner/verifier side (measured on a laptop):} \\
\multicolumn{4}{|l|}{$T_{tpmrng} = 6.3 \times 10^{-2}ms, T_{se} = 2.0 \times 10^{-3}ms, T_h = 2.1 \times 10^{-3}ms$} \\
\multicolumn{4}{|l|}{$T_{mac} =  2\times 10^{-3}ms, T_{versig} = 7 \times 10^{-2}ms, T_{ecdh} = 3.8 \times 10^{-2}ms$} \\
\multicolumn{4}{|l|}{The computation costs of operations on the device side (measured on Raspberry Pi 4):} \\
\multicolumn{4}{|l|}{$T_{rng} = 1.077 \times 10^{-2}ms, T_{sig} = 23.0569ms, T_{se} = 8.77 \times 10^{-2}ms, T_h = 2.726 \times 10^{-2}ms$} \\
\multicolumn{4}{|l|}{$T_{mac} = 2.064 \times 10^{-2}ms, T_{xor} = 2.44 \times 10^{-3}ms, T_{cp} = 1.482 \times 10^{-3}ms$} \\
\multicolumn{4}{|l|}{$T_{seal} = 1597.012ms, T_{unseal} = 533.512ms, T_{read} = 43.573ms, T_{ecdh} = 0.16ms$} \\
\hline
\end{tabular}
\end{table*}

Finally, our simulation experiment also compares with SEDA \cite{07}, WISE \cite{12}, TRAIN \cite{frolikov2025toctou}, S. Sekhavati et al. \cite{sekhavati2025resisting}, and MTRA \cite{14} in a 4-ary tree topology in the same environment. However, WISE \cite{12} does not mention how to select the devices of the specified class. We assume that the time to select devices is negligible.

\begin{figure}[!ht]
  \centering
  \begin{tikzpicture}
   
   \begin{axis}[
    width=0.9\linewidth,
    height=6cm,
    xlabel={The Number of Devices},
    ylabel={Run Time (seconds)},    
    xmin=-0.5,
    xmax=10.5,
    ymin=-3,
    ymax=75,
    xtick={0,1,2,3,4,5,6,7,8,9,10},
    ytick={0,5,10,15,20,25,30,35,40,45,50,60,70},
                ylabel style={
                    yshift=0pt
                },
    grid=both,
    clip=false,
    grid style={gray!20},
    legend style={
     rotate=1,
     at={(0.0,1.0)},
     anchor=north west
    }
    ]

    \node at (axis description cs:1.08 , 0.05) {$\times$$10^4$};
    
    \addplot[
    color=blue,
                mark=square*,
    thick
    ]
    coordinates {
     (0.1,2.469)
     (1,6.065)
     (2,8.835)
     (3,11.252)
     (4,16.632)
     (5,19.763)
     (6,22.145)
     (7,25.479)
     (8,26.871)
     (9,28.89)
     (10,32.737)
    };
    
    \addplot[
    color=green,
                mark=triangle*,
                mark size=3pt,
    thick
    ]
    coordinates {
     (0.1,14.22)
     (1,19.91)
     (2,19.93)
     (3,22.75)
     (4,22.76)
     (5,22.77)
     (6,22.78)
     (7,22.77)
     (8,22.78)
     (9,25.6)
     (10,25.59)
    };
    
    \addplot[
    color=orange,
                mark=diamond*,
                mark size=3pt,
    thick
    ]
    coordinates {
                    (0.1,0.0083462)
     (1,2.46566675)
     (2,2.56666675)
     (3,2.90470282)
     (4,3.00170282)
     (5,3.06470282)
     (6,3.17070282)
     (7,3.24070282)
     (8,3.32470282)
     (9,3.66873889)
     (10,3.74673889)
    };
    
    \addplot[
    color=pink,
                mark=asterisk,
                mark size=3pt,
    thick
    ]
    coordinates {
     (0.1,1.716)
     (1,2.439)
     (2,2.476)
     (3,2.869)
     (4,2.906)
     (5,2.881)
     (6,2.925)
     (7,2.941)
     (8,2.952)
     (9,3.347)
     (10,3.358)
    };
    
    \addplot[
    color=violet,
                mark=star,
                mark size=3pt,
    thick
    ]
    coordinates {
     (0.1,1.256)
     (1,1.813)
     (2,1.842)
     (3,2.111)
     (4,2.131)
     (5,2.152)
     (6,2.187)
     (7,2.202)
     (8,2.218)
     (9,2.430)
     (10,2.522)
    };
    
    \addplot[
    color=black,
                mark=square*,
    thick
    ]
    coordinates {
     (0.1,1.257)
     (1,1.811)
     (2,1.824)
     (3,2.080)
     (4,2.078)
     (5,2.077)
     (6,2.107)
     (7,2.106)
     (8,2.089)
     (9,2.341)
     (10,2.341)
    };
    
    \legend{ 
     S. Sekhavati et al.,
     MTRA,
     Our Scheme,
     WISE,
     SEDA,
     TRAIN
    }
    
   \end{axis}
   
  \end{tikzpicture}
  
  \caption{Performance comparisons in a 4-ary tree topology.}
  \label{fig:Perf_Compare}
  
 \end{figure}
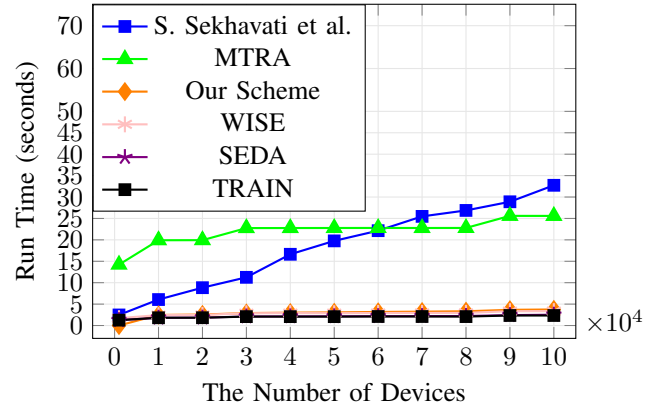

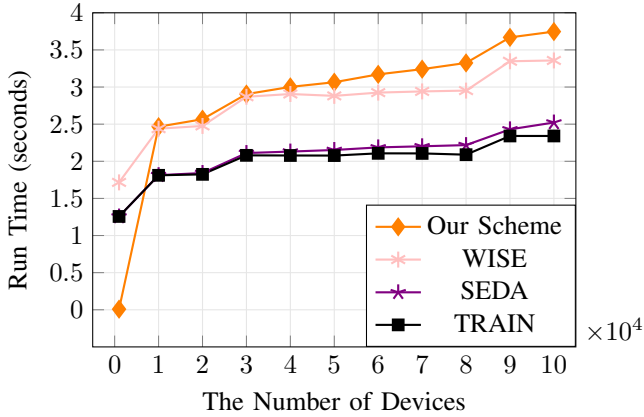
\begin{figure}[!ht]
  \centering
  
  \begin{tikzpicture}
   
   \begin{axis}[
    width=0.9\linewidth,
    height=6cm,
    xlabel={The Number of Devices},
    ylabel={Run Time (seconds)},    
    xmin=-0.5,
    xmax=10.5,
    ymin=-0.5,
    ymax=4,
    xtick={0,1,2,3,4,5,6,7,8,9,10},
    xtickmin=0,
    ytickmin=0,
    ytick distance=0.5,
                ylabel style={
                    yshift=0pt
                },
    grid=both,
    clip=false,
    grid style={gray!20},
    legend style={
     rotate=1,
     at={(1.0,0.0)},
     anchor=south east
    }
    ]

    \node at (axis description cs:1.08 , 0.05) {$\times$$10^4$};

    \addplot[
    color=orange,
                mark=diamond*,
                mark size=3pt,
    thick
    ]
    coordinates {
     (0.1,0.0083462)
     (1,2.46566675)
     (2,2.56666675)
     (3,2.90470282)
     (4,3.00170282)
     (5,3.06470282)
     (6,3.17070282)
     (7,3.24070282)
     (8,3.32470282)
     (9,3.66873889)
     (10,3.74673889)
    };
    
    \addplot[
    color=pink,
                mark=asterisk,
                mark size=3pt,
    thick
    ]
    coordinates {
     (0.1,1.716)
     (1,2.439)
     (2,2.476)
     (3,2.869)
     (4,2.906)
     (5,2.881)
     (6,2.925)
     (7,2.941)
     (8,2.952)
     (9,3.347)
     (10,3.358)
    };
    
    \addplot[
    color=violet,
                mark=star,
                mark size=3pt,
    thick
    ]
    coordinates {
     (0.1,1.256)
     (1,1.813)
     (2,1.842)
     (3,2.111)
     (4,2.131)
     (5,2.152)
     (6,2.187)
     (7,2.202)
     (8,2.218)
     (9,2.430)
     (10,2.522)
    };
    
    \addplot[
    color=black,
                mark=square*,
    thick
    ]
    coordinates {
     (0.1,1.257)
     (1,1.811)
     (2,1.824)
     (3,2.080)
     (4,2.078)
     (5,2.077)
     (6,2.107)
     (7,2.106)
     (8,2.089)
     (9,2.341)
     (10,2.341)
    };
    
    \legend{ 
     Our Scheme,
     WISE,
     SEDA,
     TRAIN
    }
    
   \end{axis}
   
  \end{tikzpicture}
  
  \caption{Detail performance comparisons of our scheme, WISE, SEDA, and TRAIN in a 4-ary tree topology.}
  \label{fig:Perf_Compare2}
  
 \end{figure}
 
Fig. \ref{fig:Perf_Compare} shows the performance comparison between our scheme and five other methods, while Fig. \ref{fig:Perf_Compare2} provides a more detailed comparison among our scheme, WISE, SEDA, and TRAIN. As shown in Fig. \ref{fig:Perf_Compare2}, our scheme is less efficient than SEDA, WISE, and TRAIN because the computational cost of TPM operation is prohibitive. Although multi-factor authentication has a linear computational cost, our method can securely protect IoT devices from passive malicious administrators attempting to access them without the owner's authentication, and it exchanges new session keys during RA to avoid attacks that exploit the time gap between successful attestation and user access to the device. 




In addition, since we adopt a tree-based structure for swarm attestation, we also evaluated the attestation time required for a single path across multiple transmission levels. The attestation time required at each level is presented in Table \ref{table:Attestation Time}. As shown in the table, attestation time increases with tree depth.

\begin{table}[!h]
\centering
\caption{Attestation Time for Hamiltonian Paths with Different Depths}
\label{table:Attestation Time}
\begin{tabular}{|c|c|c|c|c|c|}
\hline
Depths & 1 layer & 2 layer & 3 layer & 4 layer & 5 layer \\ \hline
\makecell{Attestation \\ time} & $865ms$ & $1030ms$ & $1129ms$ & $2982ms$ & $5301ms$ \\ \hline
\end{tabular}
\end{table}
\section{Conclusion}
\label{sec:Conclusion}
This work proposes a cloud-based RA system for IoT. Compared with related work, it allows session key exchange after RA. Moreover, the cloud-based IoT RA is secure against passive malicious administrators. The proposed scheme also performs comparably to related works.


%





\ifCLASSOPTIONcaptionsoff
  \newpage
\fi



%

\FloatBarrier
\bibliographystyle{ieeetr}
\bibliography{ref}
%








\end{document}